\documentclass[]{aastex7}

\usepackage{longtable}
\usepackage{rotating}
\usepackage{adjustbox}
\usepackage{lmodern}

\begin{document}
\shorttitle{Orbital planes of planetary systems by microlensing}
\shortauthors{V.Bozza and P. Rota}
\title{Orientation of orbital planes of planetary systems detected in microlensing campaigns}

\author{Valerio Bozza}
\affiliation{Dipartimento di Fisica``E. R. Caianiello''}
\affiliation{Istituto Nazionale di Fisica Nucleare, Sezione di Napoli}
\email[show]{vbozza@unisa.it} 

\author{Paolo Rota}
\affiliation{Dipartimento di Fisica``E. R. Caianiello''}
\email[show]{prota@unisa.it} 

\begin{abstract}

Galactic microlensing has the capability to determine the position angle of the detected planets in a sky reference frame. By a broad enough statistics, it is possible to investigate possible anisotropies in the distribution of the orbital planes of the planetary systems. We select 66 published microlensing planets suitable for such study and test the hypothesis that such orientations are randomly distributed against the possibility that the orbital planes follow some preferential alignment. The whole sample seems to be overall isotropically distributed, but by re-binning according to the distance along the line of sight, we find some local anisotropy peaks. Excluding those coming from very poor statistics or possible systematics, the anisotropy at 3 kpc may suggest a preferential alignment of planetary orbits in the Scutum-Centaurus spiral arm of the Milky Way with the Galactic plane. Special orientations of the orbital planes may be reminiscent of the specific conditions that triggered and drove the star formation processes and how these are related to local and global Galactic kinematics. Using the method proposed here, the future Roman microlensing survey will be able to identify and quantify preferential orientations in all structures from the Sun to the bulge with high confidence and accuracy.
\end{abstract}



\section{Introduction}

Before the discovery of extrasolar planets, the hypothesis of the existence of preferential orientations of the orbital planes has been posed for stellar binary systems. Earlier studies based on eclipsing binaries have found that the fraction of stars undergoing eclipses is similar all over the sky \citep{Huang1966}. If there were a preferential orientation along the Milky Way plane, we would have a higher frequency of eclipsing binaries in the Galactic plane than toward the Galactic poles. For spectroscopic binaries, the inclination of the orbital plane is unknown, but it is possible to determine the line of the apsida \citep{Brazhnikova1975}. 

For visual stellar binaries, it is possible to accurately determine the orbital planes and directly investigate the distribution of the directions of the angular momenta in the sky. If we take a sufficiently large sample, indeed the distribution is fairly isotropic, but some possible alignment arises at scales below 10 pc \citep{Glebocki2000}. This ``anomaly'' has been confirmed by later studies \citep{Kisselev2009}, which have found a particular accumulation of the angular momenta along the direction at Galactic coordinates $l = 46.0^\circ, b = 37^\circ$ for stars closer than 8.1 pc \citep{Agati2015}. This direction is $41^\circ$ apart from the ecliptic pole, a fact that may be suggest that even our Solar System may be marginally included in this accumulation. 

Unfortunately, visual binaries are limited to the solar neighborhood. If we want to study the orientations of binary systems on larger scales, we may resort to extended objects associated with binary systems. Careful studies of asymmetric planetary nebulae in our Galaxy find that they are mostly isotropic, with the somewhat unexpected result that binary systems in the Galactic bulge have orbital planes preferentially orthogonal to the Galactic plane \citep{Melnick1975,Weidmann2008,Rees2013,Ritter2020,Tan2023}. Finally, gravitational waves in mHz have been suggested to test any alignment of binary systems in the Galaxy \citep{Seto2024}.

For the orientation of the orbital planes of planetary systems there are no similar investigations and it is generally assumed that orbital planes are randomly and isotropically distributed, whatever the chosen line of sight. Only a recently found deficit in transiting planets for high-galactic-amplitude stars could be also explained by some preferential alignment \citep{Zink2023}. In principle, the orbital elements of directly imaged planets can be determined over a sufficiently long observation time. This has been done in a number of cases and can pose the basis for future statistics of systems in the Solar neighborhood \citep{Ferrer2021,Rice2024,Maire2023}. Similarly, circumstellar disks observed in the infrared are growing in number and may serve the same purpose \citep{Andrews2020}. Statistics on orbital planes would have important implications under many aspects related to the formation of planetary systems after the collapse of a protostellar cloud, in connection with the dynamics of turbulent fragmentation and the role of magnetic fields \citep{Matsumoto1997,Orkisz2017,Krumholz2019}. 

The existence of any preferential orientations of the orbital planes would pose a severe problem for frequency estimations of extrasolar planets. In fact, such estimates implicitly assume that planetary planes are randomly distributed. It is therefore very important to validate (or falsify) such assumptions: any deviations from this rule would appear as unaccounted biases in exoplanetary demographics \citep{Demo2022}.

In this paper, we propose a new method to investigate the orientation of orbital planes of the planets discovered by the microlensing method. The great advantage of the microlensing planets with respect to those discovered by other methods is that they lie all along the line of sight to the Galactic bulge, thus enabling a detailed study of the orientations in different regions of the Galactic plane, from our local spiral arm to the bulge, passing across at least two well-established spiral arms (the Carina-Sagittarius and Scutum-Centaurus arms). Any possible local preferential orientations may thus be connected with these morphological macro-structures. 


In brief, microlensing occurs when a planetary system passes close to the line of sight to a distant background star \citep{Gaudi2012}. The planetary system acts as a gravitational lens, temporarily magnifying the flux of the background star. From the analysis of the photometric light curve, it is possible to derive several parameters of the planetary system, such as the mass ratio and the projected separation. Most important to our study is the angle $\alpha$ between the proper motion $\mathbf{\mu}_{rel}$ of the lens relative to the source and the lens-planet axis. This is very firmly established by the shape of the light curve and is measured with negligible error in all microlensing planets. The orientation of the relative proper motion in the sky can be obtained if the parallax effect is measured \citep{Gould1992,Gould2000}. This appears as a long-term modulation due to the Earth orbital motion around the Sun (annual parallax). In alternative, the parallax can be measured by observing with a spacecraft at $\sim$au separation, as successfully achieved by {\it Spitzer} in a number of cases \citep{Udalski2015,Zhu2017}. Finally, the relative proper motion and its direction can be directly measured with high-resolution imaging follow-up a few years after the microlensing event, when lens and source are resolved as separate objects \citep{Batista2015,Bennett2015}. By simple combination of these two pieces of information, we obtain the position angle of the planet in the sky with respect to its host at the peak time of the microlensing event. 

The paper is organized as follows. In Section 2 we introduce the statistical sample used for our study and the basic concept of the measurement in detail. Section 3 derives the distribution of expected position angles from a parameterized anisotropic distribution of orbital planes. In Section 4 we introduce the likelihood function used to compare observations with expectations and discuss possible sources of bias. The comparison with the observations and the results are discussed in Section 5. Section 6 contains the conclusions.

\section{Concept and statistical sample}

\begin{figure}[t]
 \centering
 \includegraphics[width=8cm]{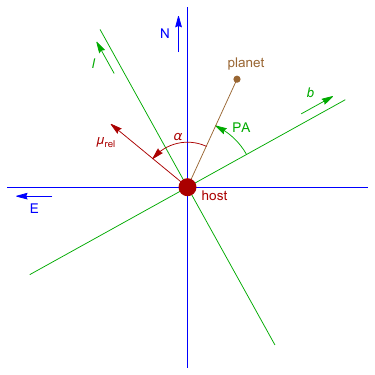} 
 \caption{Geometric determination of the planet position angle $PA$ in a microlensing event. The angle $\alpha$ between the host-planet direction and the proper motion $\mathbf{\mu}_{rel}$ of the lens relative to the source is very precisely constrained from the modeling of the microlensing light curve. The North-East components of the proper motion are determined by the parallax effect or by high-resolution imaging. The blue axes represent the North and East directions, while the green axes represent the directions for increasing Galactic latitude $b$ and longitude $l$.}
 \label{Fig geometry}
 \end{figure}

As illustrated by Fig. \ref{Fig geometry}, the position angle of the planet relative to the Galactic pole direction is
\begin{equation}
    PA=\widehat{GN} + \arctan\frac{\mu_{rel,E}}{\mu_{rel,N}}-\alpha, \label{PostionAngle}
\end{equation}
where $\widehat{GN}$ is the angle between the direction to the Galactic pole and the North direction. This angle depends on the specific coordinates of the microlensing event in the sky. For currently known microlensing planets detected toward the bulge fields, this angle is always about $61^\circ$. The components of the relative proper motion of the lens with respect to the source in an equatorial frame are $(\mu_{rel,N},\mu_{rel,E})$. As anticipated in the introduction, $\alpha$ is the angle between the relative proper motion and the star-planet axis.

By careful inspection of all published microlensing planets, we have found that among the 235 planets reported to date in the NASA exoplanet archive\footnote{\url{https://exoplanetarchive.ipac.caltech.edu/}}, 66 planets have a measure of the parallax components in the proposed models and are thus eligible to be included in our study. In general, the inclusion of annual parallax in the model always leads to some improvement, but only when this improvement is significant enough and is evident in the residuals it is taken in consideration in the physical interpretation of planetary microlensing events. We have thus considered only those events in which the parallax measurement was reported and included in the physical discussion. Some of them have very accurate parallaxes, especially those featuring a space observation, but there are others with very poor detections, resulting in large error bars for the position angle of the planet. Furthermore, a good fraction of the events have double or multiple degeneracies.  Finally, a few microlensing events have led to the discovery of two planets in the same system. In principle, for some microlensing events, some rough information on the orbital motion is available, but it is too limited to be considered in our study.
The full list of planets included in our sample is available in Appendix \ref{Sec Appendix sample}. In Table 2 we collect planets with annual parallax or satellite parallax (marked with a dagger). In Table 3 we have planets whose proper motion is known by High-Resolution Imaging measurements.

\begin{figure}[t]
 \centering
 \includegraphics[width=11.5cm]{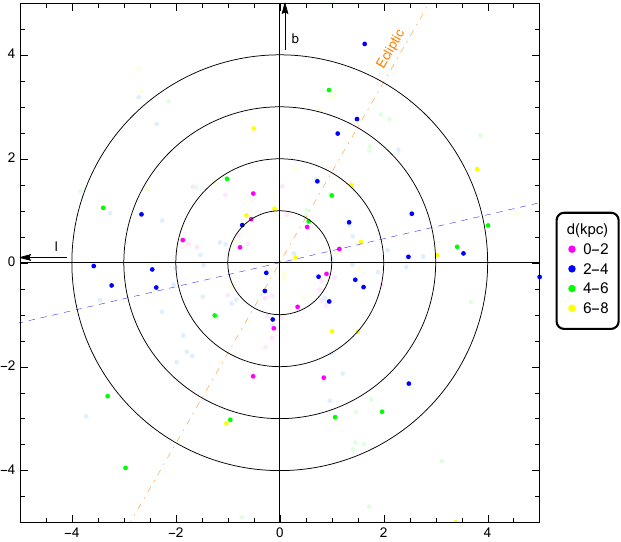} 
 \caption{Relative position in the sky of the microlensing planets with respect to their hosts. Different colors distinguish planets in different distance intervals from the Solar System; the best models are shown with full color, while alternative disfavored models, when present, are shown in a more transparent color. The directions of increasing Galactic latitude $b$ and longitude $l$ are marked, together with the ecliptic plane. The dashed colorful line marks the preferential plane we find for planets in the $2-4$ kpc range (blue). Circles mark projected separations from the host in units of au. }
 \label{Fig scatterplanets}
 \end{figure}
 
Fig. \ref{Fig scatterplanets} shows the locations of the planets relative to their host stars. Taken as a single population, microlensing planets do not seem to show any preferential orientations. However, if we distinguish planets according to their distance from the Solar System, at least one preferential orientation seems to arise, as marked by the dashed blue line in the figure, averaging on planets located between 2 and 4 kpc from the Sun. In order to put this impression on a solid statistical basis, we need to describe the probability to find a planet at some position angle, given the orbital inclination with respect to the line of sight. While an edge-on planetary system would only yield two possible position angles for the planet, a face-on system would yield any position angles with the same probability. In-between these two extreme cases, we have intermediate inclinations, which would result in continuously varying probability distributions, always favoring those angles close to the line of nodes.

\section{The distribution of position angles} \label{appB}

\subsection{A simple ansatz for an anisotropic distribution}

Now, let us assume that a preferential orientation $\hat{R}$ exists for the orbital angular momenta of the planets. If we denote by $i$ the inclination of the angular momentum of an individual planet with respect to the preferred direction $\hat R$, an isotropic distribution would be uniform in $\mathrm{d}\cos i$. We parameterize a generic deviation from isotropy by the von Mises-Fisher distribution, which is commonly used in anisotropy problems

\begin{equation}
    \frac{\mathrm{d}N}{\mathrm{d} \cos i}= \frac{q}{4\pi\sinh q}\exp(q \cos i) \label{pcosi}
\end{equation}

\begin{figure}[t]
 \centering
 \includegraphics[width=8cm]{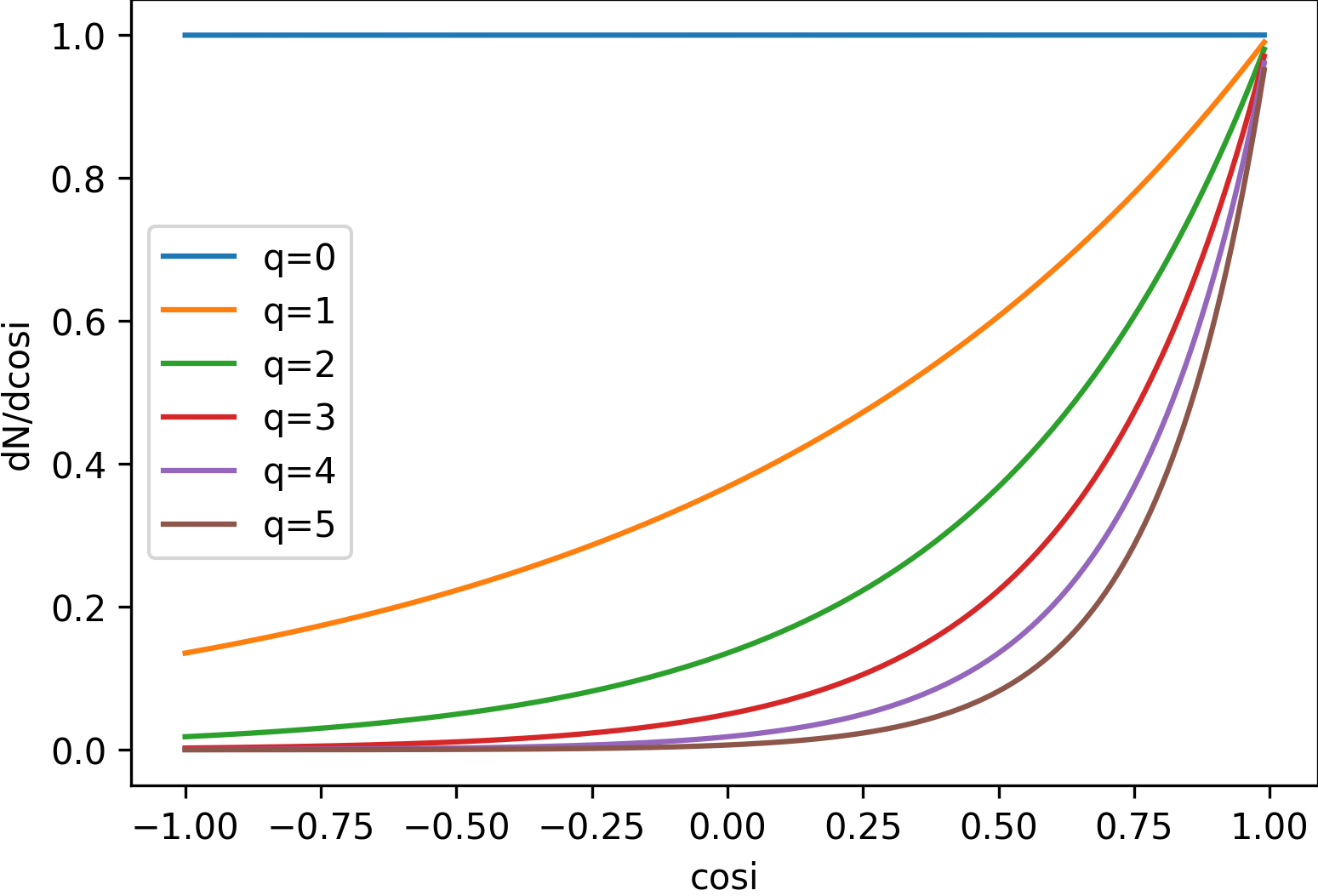} 
 \caption{Assumed distributions for the orbital angular momenta of individual planets with respect to the reference direction $\hat R$ for different values of $q$: $q=0$ corresponds to an isotropic distribution. For visualization purposes, distributions have been normalized to one at $\cos i=1$.}
 \label{Fig dNdcosi}
 \end{figure}

If the anisotropy parameter $q$ vanishes, we recover an isotropic distribution independent of $\cos i$. As $q$ grows, the distribution favors inclinations close to $i=0$ more and more. Such ansatz is representative of any reasonable intermediate situations between the isotropic $q=0$ and the fully aligned $q\rightarrow \infty$ extreme cases. Fig. \ref{Fig dNdcosi} shows the von Mises-Fisher distribution for different values of $q$.

\subsection{Expected distribution of position angles}

Starting from the distribution (\ref{pcosi}), it is possible to derive the expected distribution of the position angles of the planets in the sky. We expect a flat distribution for $q=0$ and a distribution peaked around the line of nodes for large values of $q$. For simplicity, we assume that $\hat R$ is orthogonal to the line of sight to the Galactic center $\hat z$, so that $\hat R \cdot \hat z=0$. We will come back on this assumption at the end of our derivation. The direction $\hat x= \hat R \times \hat z$ completes an orthonormal basis in which the plane of the sky is spanned by ($\hat x , \hat R)$. 

The vector $\hat R$ will have its position angle in the sky with respect to the Galactic pole $\widehat{GN}$, denoted by $PA_R$. Therefore, we may express the orientations of planets in the sky with respect to this reference axis as
\begin{equation}
    PA'=PA-PA_R
\end{equation}
and then investigate whether any anisotropies arise in the distribution of $PA'$.

Each individual planet will have its own orbital angular momentum $\hat L$ with inclination $i$ from the reference direction $\hat R$. Once we fix the inclination $i$, the angular momentum direction will be fully specified by another angle $\Omega$ that we define as the angle between the sky plane $(\hat x, \hat R)$ and the plane $(\hat R,\hat L)$ containing the reference direction and the angular momentum. Finally, the planet will be found at some anomaly $\varphi$ along its orbit. We take this angle starting from the intersection of the orbital plane and the plane $(\hat x,\hat z)$ orthogonal to the reference direction. 

It is fair to assume that planets will be uniformly distributed along their orbit. So, the probability should  not depend on $\varphi$. Similarly, we assume that orbits with the same inclination $i$ with respect to the reference direction will be equally probable, with no dependence on the node line location $\Omega$. Therefore, we only assume that orbits will be generated with a probability depending on the inclination $i$ from the reference direction $\hat R$ as specified in our ansatz (\ref{pcosi}). We do not expect to have the sensitivity to distinguish between our ansatz and any alternative distributions, but we are confident that any gross anisotropies will favor a high q distribution over the isotropic one.

Therefore, the probability to find a planet with angular momentum at inclination $i$ oriented along $\Omega$ and anomaly $\varphi$ will be
\begin{equation}
    \mathrm{d}N= \frac{\mathrm{d}N}{\mathrm{d}\cos i} \mathrm{d}\cos i \; \mathrm{d} \Omega \; \mathrm{d} \varphi, \label{ptot}
\end{equation}
where the only non-trivial dependence is on $\cos i$ through Eq. (\ref{pcosi}).

As $\varphi$ spans all values between $0$ and $2\pi$, the position angle of the planet in the observer's sky will take all values in $[0,2\pi]$. The two variables can be interchanged once we find the correct relation, which is
\begin{equation}
    \tan \varphi= \frac{\cos PA' \sin \Omega}{\cos PA' \cos i \cos \Omega +\sin PA' \sin i}. \label{varphi_PA'}
\end{equation}

We can thus change the variable $\varphi$ with $PA'$ in Eq. (\ref{ptot}) and get
\begin{equation}
    \mathrm{d}N= \frac{\mathrm{d}N}{\mathrm{d}\cos i} \frac{\mathrm{d}\varphi}{\mathrm{d}PA'} \mathrm{d}\cos i \; \mathrm{d} \Omega \; \mathrm{d} PA', \label{ptotmod}
\end{equation}
where the derivative $\frac{\mathrm{d}\varphi}{\mathrm{d}PA'}$ can be explicitly calculated from Eq. (\ref{varphi_PA'}).

\begin{figure}[t]
 \centering
 \includegraphics[width=8cm]{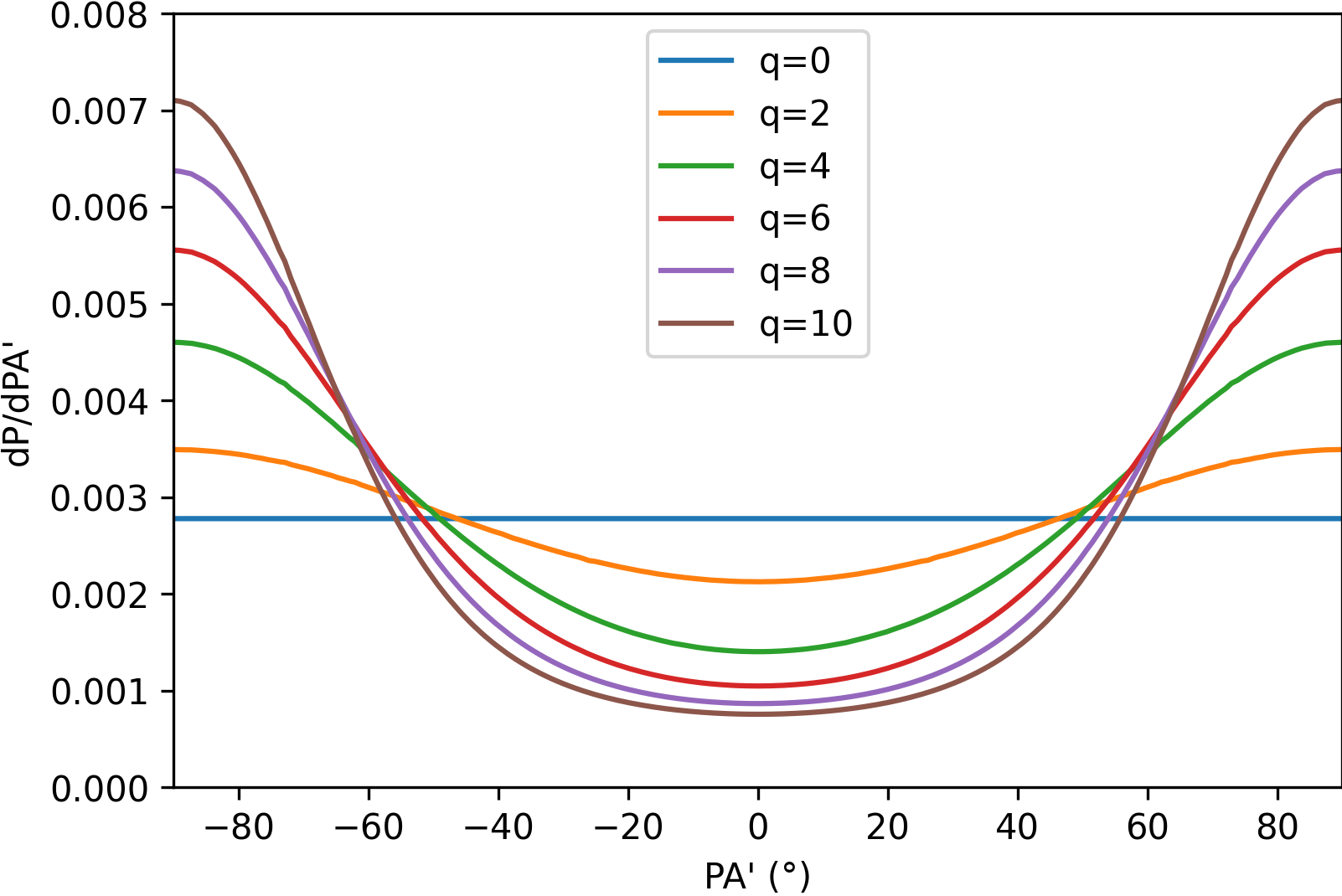} 
 \caption{Normalized probability distribution for the position angle $PA'$ of planets with respect to the reference direction $\hat R$ for different values of the anisotropy parameters $q$.}
 \label{Fig pPA}
 \end{figure}

At this point, in order to obtain the distribution of planets in position angle $PA'$, we just have to integrate over the other variables $\Omega$ and $\cos i$
\begin{equation}
    \frac{\mathrm{d}N}{\mathrm{d}PA'}= \int\limits_{-1}^{+1}\mathrm{d}\cos i \int\limits_0^{2\pi} \mathrm{d} \Omega\frac{\mathrm{d}N}{\mathrm{d}\cos i} \frac{\mathrm{d}\varphi}{\mathrm{d}PA'} . \label{dNdPA}
\end{equation}
This can be easily done numerically for any values of $q$ at any given position angle $PA'$. Fig. \ref{Fig pPA} shows the normalized probability as a function of $PA'$ for different values of the anisotropy parameter $q$ as obtained here.

The distribution of the planet position angles is flat for $q=0$. As $q$ grows, Planets accumulate in the plane orthogonal to $\hat R$ corresponding to $PA'=\pm 90^\circ$.

In our analysis we have assumed that the preferential direction of the angular momenta $\hat R$ is orthogonal to the line of sight. It is possible to relax this restriction and let $\hat R$ form an angle $\psi$ with the plane of the sky. Even with this additional parameter it is possible to carry the full calculation. However, we simply note that the distributions become flatter and flatter as $\psi$ approaches $\pi/2$. As the preferential direction for the angular momenta $\hat R$ aligns with the line of sight, orbital planes will always be seen face-on and planets will be seen at all position angles indifferently. Assuming some parameters $(q,\psi)$ for the true distribution, we can always find a $\hat q$, such that the distribution with $(\hat q,0)$ mimics the true distribution almost perfectly, with deviations of the order of few percent. Since the available statistics would not be sufficient to distinguish such similar distributions and break the degeneracy, in our analysis we have just assumed a reference direction $\hat R$ orthogonal to the line of sight and tracked the anisotropy by the single parameter $q$. 


\section{Building the likelihood function}

The distributions in Fig. \ref{Fig pPA} represent our expectations for the position angles of the planets for different anisotropy levels. In order to check the (an-)isotropy of the microlensing planets, we should compare the distribution of the observed position angles in Fig. \ref{Fig scatterplanets} with the theoretical expectations in Fig. \ref{Fig pPA} for different position angles of the reference direction $PA_R$ and different anisotropy levels $q$. This can be done by evaluating the likelihood for our theoretical distributions $\frac{\mathrm{d}N}{\mathrm{d} PA'}$ as a function of the parameters $PA_R$ and $q$. 

The likelihood is just the product of the probabilities to find our planets in the positions in which we observe them, given the assumed theoretical distribution. 
However, since the observed values are affected by uncertainties that can be very large in some cases, we replace the best values of the position angles retrieved by microlensing modeling with gaussian functions whose width is established by the uncertainty in the position angles. Therefore, we calculate the likelihood by averaging over these gaussian functions as follows
\begin{equation}
    \mathcal L (PA_R,q) = \exp \int \mathrm{d} (PA) \; \log \left[\frac{\mathrm{d}N}{\mathrm{d} PA'}(PA';PA_R,q) \right] \sum\limits_{i,j} w_{ij} g\left(\frac{PA-PA_{ij}}{\sigma_{ij}}\right), \label{likelihood}
\end{equation}
where the summation index $i$ runs over the planets in the sample, $j$ runs over the alternative models with a corresponding weight factor $w_{ij}$. $PA_{ij}$ and $\sigma_{ij}$ are the position angle and uncertainty for model $j$ of planet $i$ and $g(x)$ is a normalized gaussian function. The integral runs over all position angles $PA$, but the distribution depends on the position angle $PA'=PA-PA_R$ relative to the reference direction.

It is common in microlensing to have some discrete degeneracies that remain at the end of the analysis, such as the close/wide or the offset degeneracy \citep{Zhang2022}. In these cases, we include all alternative solutions weighed by a factor $\tilde w_{ij}\propto e^{-\chi^2_{ij}/2}$ such that $\sum\limits_j \tilde w_{ij}=1$. Besides discrete degeneracies, the full $w_{ij}$ appearing in Eq. (\ref{likelihood}) can also include further factors accounting for any observational biases (see \ref{Sec bias}). 

We also have some multi-planetary systems. It is reasonable to assume that the two planets have similar orbital planes. Since we are building a statistics on the orbital planes of planetary systems as a whole, two-planet systems contribute as two independent measurements on the same system. Therefore, we use both measurements halving their weights (see Eq. \ref{weight}) so as to avoid double-counting the same system. 

We note that if we replace the gaussians by delta functions the expression (\ref{likelihood}) collapses to the product of probabilities $\frac{dN}{dPA'}$ evaluated at the planet position angles. The likelihood so defined is a function only depending on the reference direction position angle $PA_R$ and the anisotropy parameter $q$.

\subsection{Accounting for biases from microlensing efficiency} \label{Sec bias}

\begin{figure}[t]
\centering
\includegraphics[width=8cm]{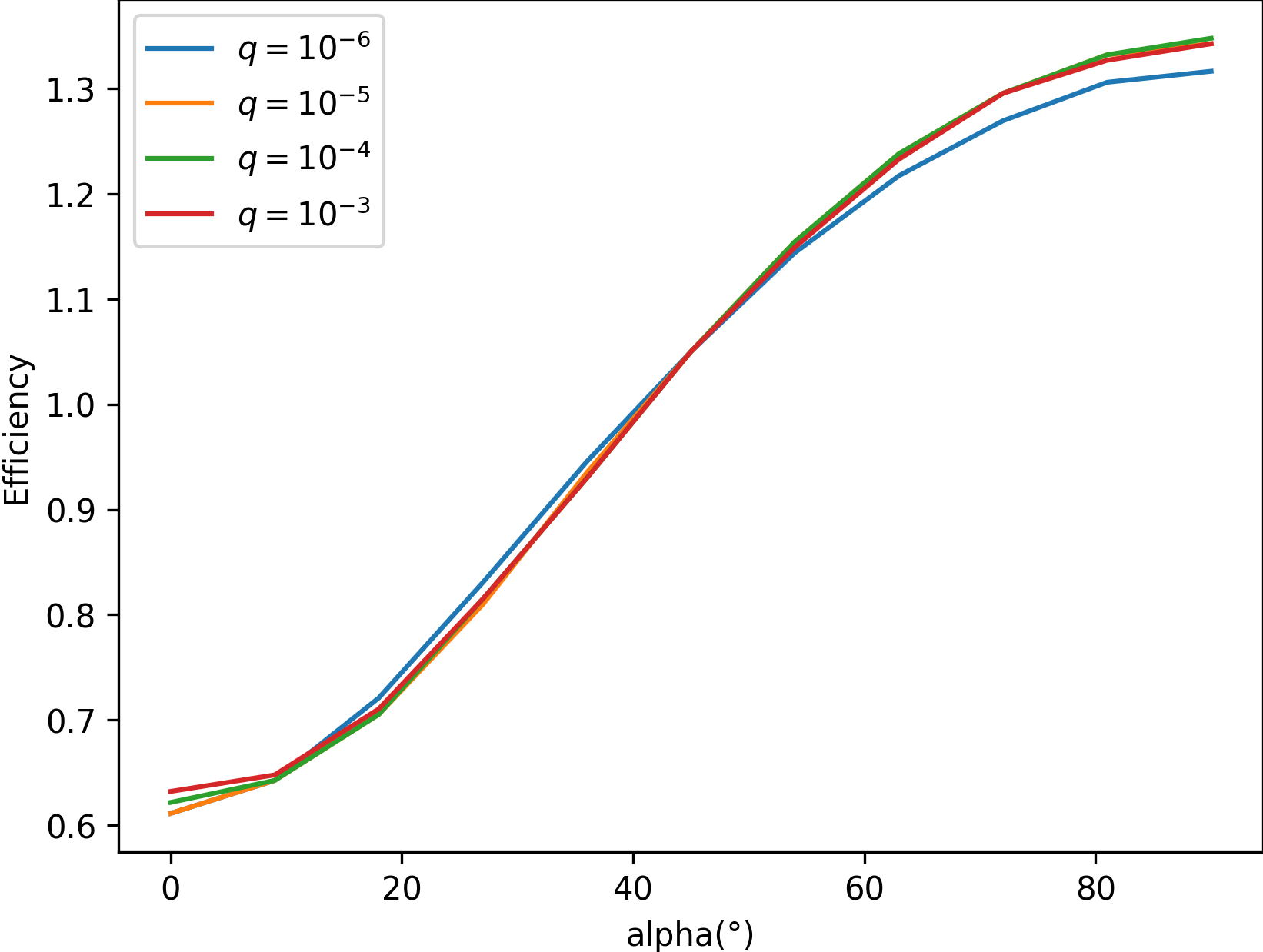} 
\caption{Example of microlensing efficiency for planet detection as a function of the angle $\alpha$ between the planet-lens axis and the relative proper motion between lens and source. This plot is calculated for $s=0.8$}
\label{Fig eff alpha}
\end{figure}

In this study we are using the position angles of the microlensing planets to build up a statistics to be compared with our theoretical distribution. However, microlensing may be biased toward some specific position angles by unaccounted selection effects. It is in fact well known that the distribution of relative proper motion between lens and source is asymmetric due to the rotational motion of the disk (see e.g. Fig. 2 of \citet{Koshimoto2020}). This would not be a problem if the angles $\alpha$ between the proper motion direction and the planet-star direction were truly randomly distributed. In reality, the microlensing efficiency depends on $\alpha$, as it can be appreciated from Fig. \ref{Fig eff alpha}\footnote{A similar plot can be found in \citep{Saggese2025}, while typical efficiency calculations average over $\alpha$ and focus on the dependencies on planetary mass and separation \citep{Gaudi2000}.}. We have derived it by counting the number of trajectories featuring a planetary deviation with $\Delta \chi^2$ above some threshold for a planet at given separation $s$ and mass ratio $q$ for a trajectory with angle $\alpha$ as we vary the impact parameter $u_0$. Such ``cross section'' has then been normalized with respect to the average over all $\alpha$'s, thus removing the dependence on the chosen threshold. 

Since each planet $i$ in our sample has its own parameters, we keep all parameters from each model $j$, except $\alpha$ and $u_0$, and calculate the normalized cross sections $\epsilon_{ij}(\alpha)$ as in Fig. \ref{Fig eff alpha}. The weight assigned to the model of the planet in the likelihood (\ref{likelihood}) is then divided by such efficiency. Taking into account the multiplicity of the planetary system $m$, the full weight of each model $j$ for planet $i$ is
\begin{equation}
    w_{ij}= \frac{\tilde w_{ij}}{\epsilon(\alpha_{ij})m}. \label{weight}
\end{equation} 

A planet found in spite of a lower efficiency will thus count a bit more than a planet found at an $\alpha$ with higher efficiency. By restoring a flat distribution in $\alpha$, we compensate for the only clear bias effect intervening in our analysis. We note that this correction is relatively mild, since the efficiencies for different angles change by a factor of two at most.

Further comments on other possible sources of bias are given in the conclusions.

\section{Results}

\begin{figure}[t]
 \centering
 \includegraphics[width=12cm]{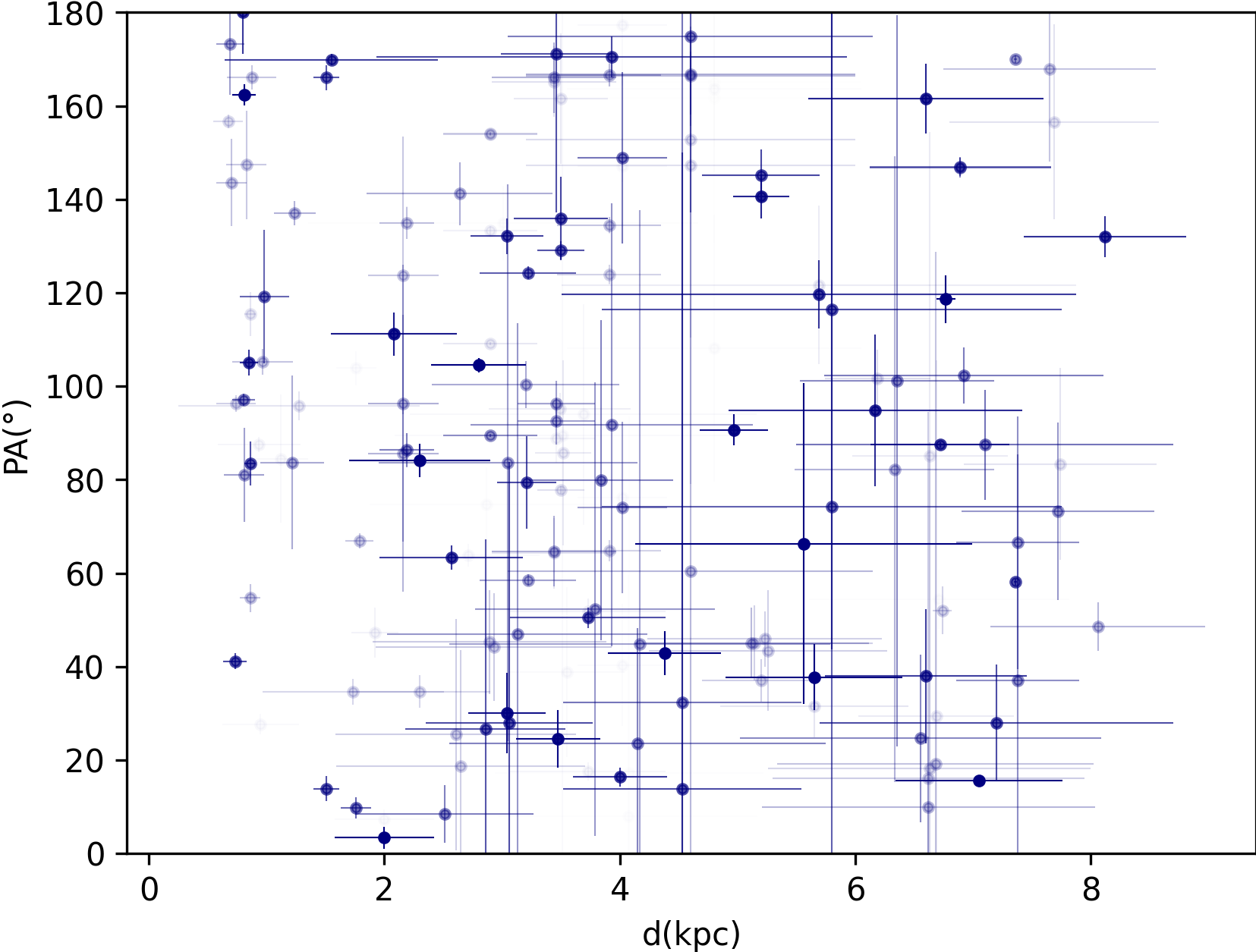} 
 \caption{Planets considered in this study. In the horizontal axis we have the distance from the Solar System. In the vertical axis we have the position angle from Galactic North. Since only the orbital plane of the planet matters, all angles $PA>180^\circ$ are reported as $PA-180^\circ$. For planets for which degenerate models are available, we report the position angles derived by all models with a transparency indicating the relative weight, calculated as explained in Eq. (\ref{weight}).}
 \label{Fig PA vs DOL}
 \end{figure}

Microlensing planets in our sample lie all along the line of sight to the Galactic bulge, thus belonging to different morphological and dynamical components, possibly including the bar and different spiral arms. Fig. \ref{Fig PA vs DOL} shows how microlensing planets in our study are distributed in a plane formed by distance and position angle from the Galactic North. Accumulations of planets at some position angle $PA$ would indicate an anisotropy with $PA_R = PA \pm 90^\circ$. We also note that by symmetry we only let $PA_R$ vary in the range $[-90^\circ, +90^\circ]$.

\begin{figure}[t]
 \centering
 \includegraphics[width=\linewidth]{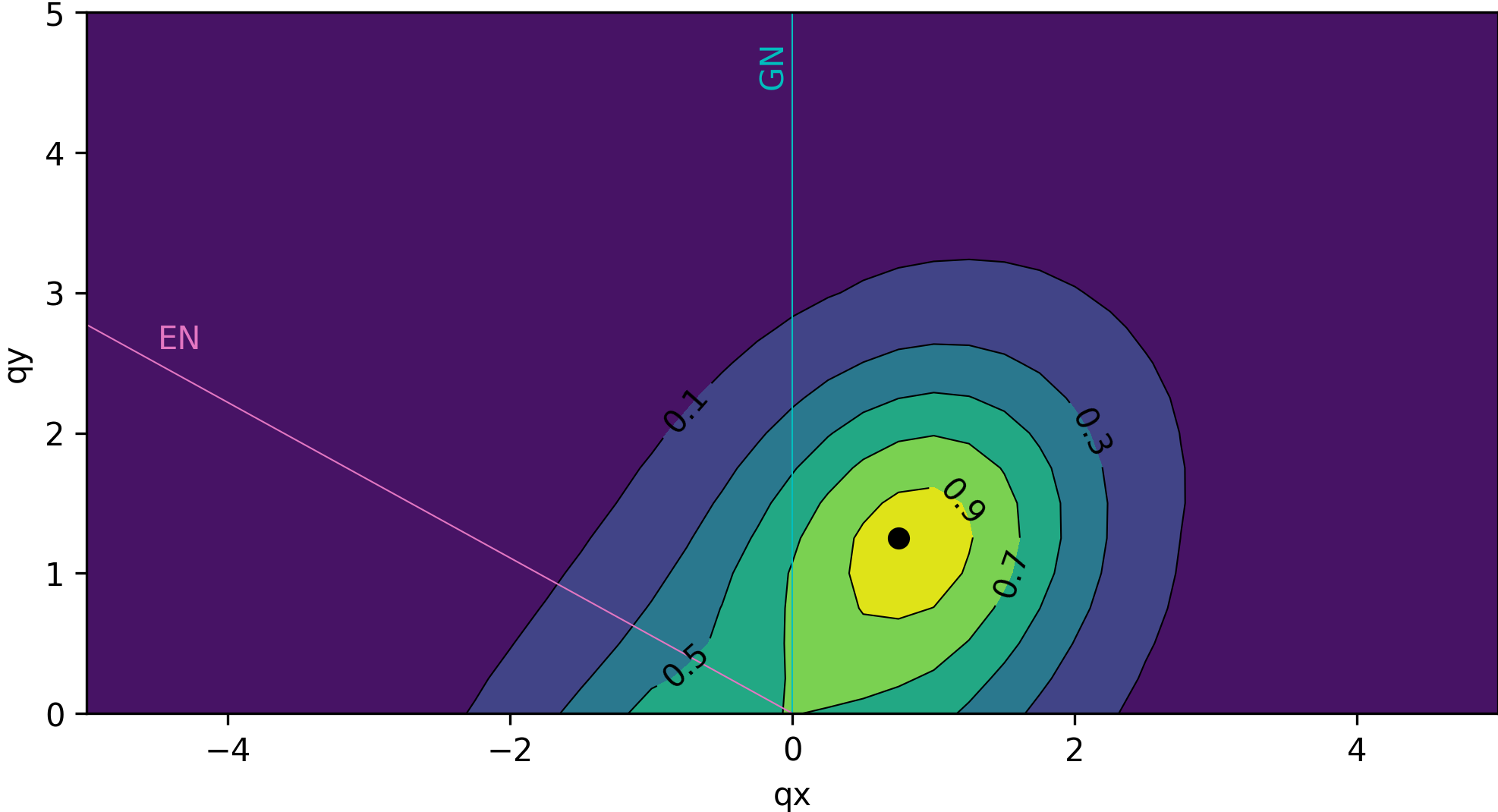} 
 \caption{Likelihood contours in the parameter space $(q_x,q_y)$ for the whole sample of microlensing planets. The likelihood is normalized to 1 at the peak position, marked by a dot. We remind that an isotropic distribution corresponds to $q_x=q_y=0$. The GN line indicates the direction of Galactic North, while EN indicates the Ecliptic north.}
 \label{Fig likelihood}
 \end{figure}

Indeed, if we take our whole sample of microlensing events, the likelihood analysis shows that an isotropic distribution perfectly fits the data, as shown in Fig. \ref{Fig likelihood} in the plane $(q_x,q_y)\equiv (q \cos PA_R, q \sin PA_R)$. The likelihood at $q=0$ is just 0.68 the maximum likelihood, which is reached at $q=1.45$, $PA_R=-31^\circ$. Such slight preference for an anisotropic model is quite insignificant and indeed a fully isotropic model is well within all contours up to 0.9 $\mathcal{L}_{max}$. In practice, microlensing planets taken altogether give a strong support to the simple assumption of randomly distributed orbits.

\begin{figure}[t]
 \centering
 \includegraphics[width=12cm]{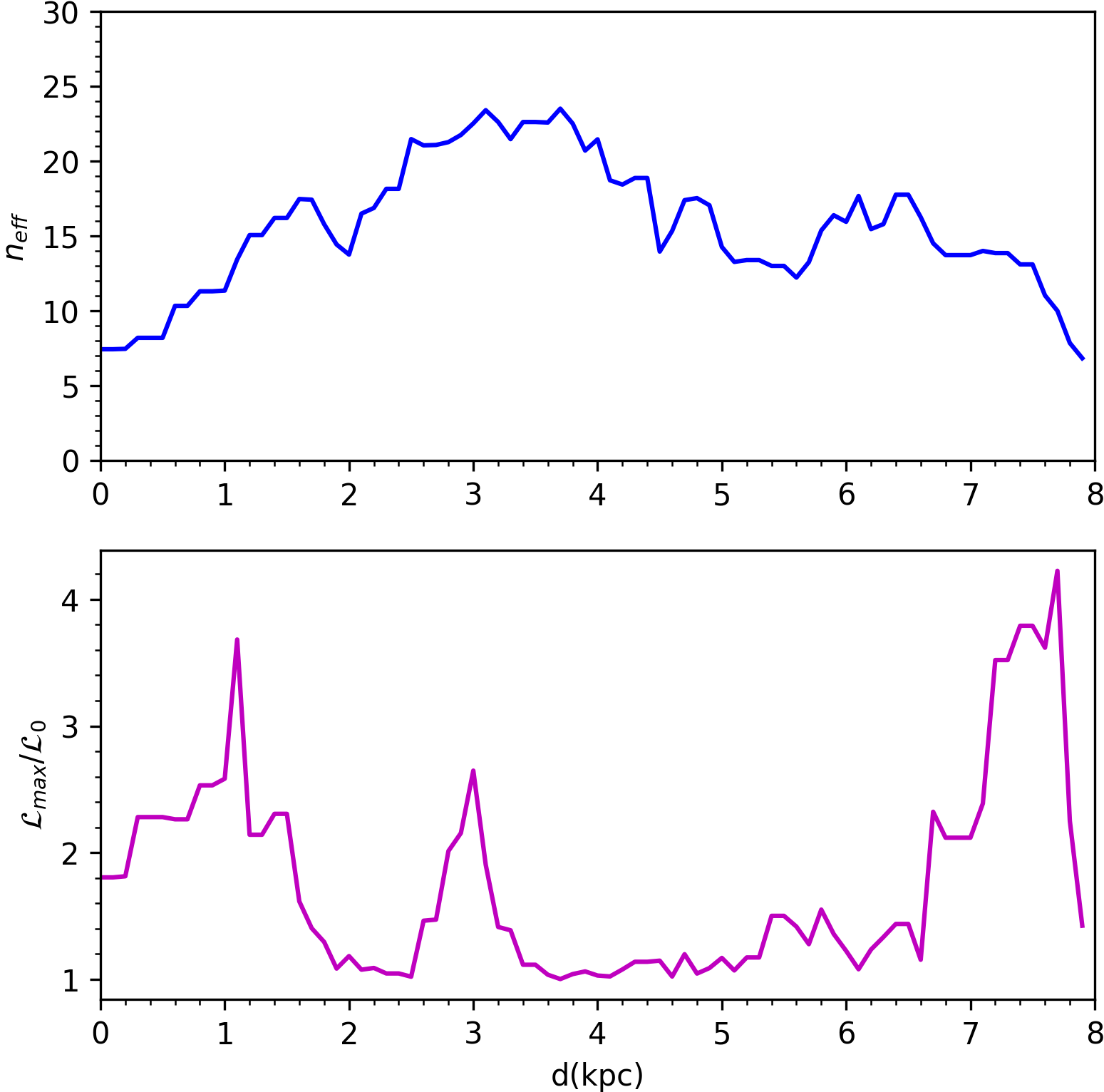} 
 \caption{Top panel: effective number of planets in a window 2 kpc wide centered at distance $d$. These are the planets used to construct the following figures. Bottom panel: ratio of the maximum likelihood to the likelihood for an isotropic distribution.}
 \label{Fig n L vs DOL}
 \end{figure}

However, we can probe the preferential orientations of planetary orbits in different regions of our Galaxy from our local arm up to the bulge by selecting planets at particular distances. To this purpose,  we study of the anisotropy likelihood as a function of the planet distance $d$ from the Solar System by only keeping planets within 1 kpc from $d$, thus defining a moving window of size 2 kpc to harvest sufficient statistics for our analysis. The top panel of Fig. \ref{Fig n L vs DOL} shows the effective number of planets falling in such moving window as a function of the distance from the Solar System. Note that the weight of each planet is corrected by the efficiency as from Eq. (\ref{weight}), so we have $n_{eff}=\sum\limits_{i,j}w_{ij}$. The number of planets is smaller close to the observer and in the bulge, while most of the planets with parallax measurements fall at mid-range. 

In the bottom panel of Fig. \ref{Fig n L vs DOL} we show the main result of the likelihood calculation, namely the ratio $\mathcal{L}_{max}/\mathcal{L}_0$ between the maximum likelihood and the likelihood for the isotropic distribution. The higher this ratio, the stronger the evidence for anisotropies. We can see that the value is generally very small and close to 1, indicating that the distribution is fairly isotropic in most regions of the Galaxy. However, there are three distinct peaks at 1.1, 3 and 7.7 kpc that draw our attention.

\begin{figure}[t]
 \centering
 \includegraphics[width=12cm]{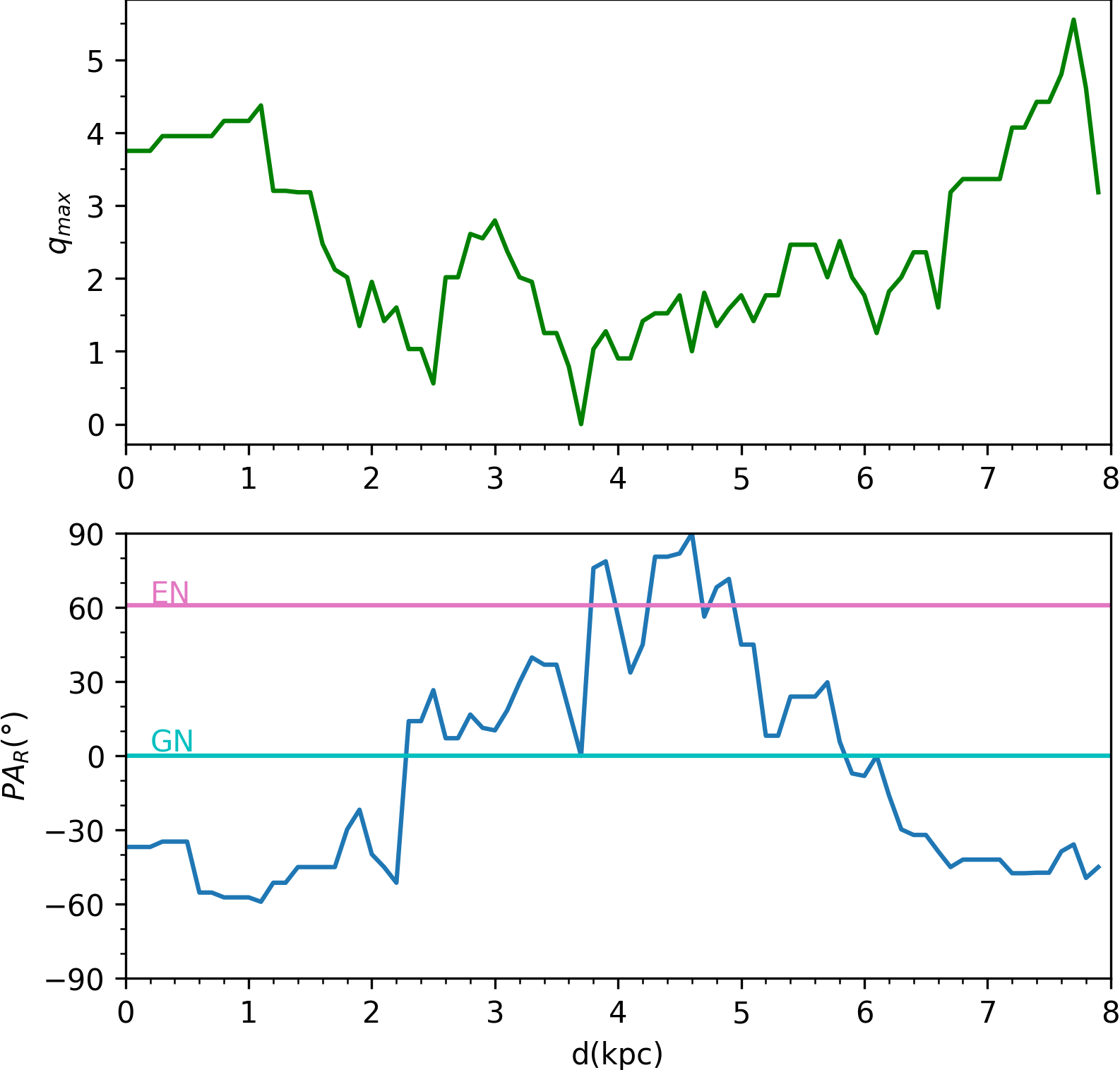} 
 \caption{Top panel: Most likely anisotropy parameter $q$ as a function of the distance from the Solar System. Bottom panel: Most likely position angle $PA_R$ of the preferential direction of angular momenta.}
 \label{Fig q PAR vs DOL}
 \end{figure}

Fig. \ref{Fig q PAR vs DOL} shows the preferred values of the anisotropy parameter $q$ and the position angle $PA_R$ of the reference direction of the angular momenta. Indeed, for all three peaks in the likelihood ratio we find a distinct peak in the anisotropy $q$. The characteristics of the three peaks are summarized in Table \ref{Tab results}. The peaks at $d=1.1$ and $d=7.7$ kpc suggest anisotropies close to the Solar neighborhood and in the bulge. However, they are derived from a relatively low number of planets (13.4 and 10 respectively), exposing them to the concern that they may be the outcome of some random fluctuations. In addition, it is quite suspicious that they come at the two extrema of the distance range: unidentified biases in the observations of microlensing planets might have produced this result. Indeed, both peaks lie at negative position angles, almost perpendicular to the ecliptic north. This means that microlensing planets are more easily found (or have measurable parallax) in a plane orthogonal to to the ecliptic. Although we have corrected for the efficiency in $\alpha$, there might be an additional selection bias that needs further investigation.
 
\begin{table}
\begin{tabular}{|l|cccc|}
\hline
Peak position (kpc) & $q$ & $PA_R$ &  $\mathcal{L}_{max}/\mathcal{L}_0$ & $n_{eff}$\\
\hline
$1.1$ & $4.4$ & $-59.0^{\circ}$ & $3.7$ & 13.4 \\
$3.0$ & $2.8$ & $10.3^{\circ}$ & $2.6$ & 22.5 \\
$7.7$ & $5.6$ & $-35.8^\circ$ & $4.2$ & 10.0\\
\hline
\end{tabular}  \label{Tab results} \caption{Characteristics of the three local anisotropy peaks found in the orientations of microlensing planets: peak position from the Solar System, anisotropy parameter $q$, preferred position angle $PA_R$ of the angular momenta from the Galactic North, likelihood ratio $\mathcal{L}_{max}/\mathcal{L}_0$ to the isotropic distribution, effective number of planets $n_{eff}$ used for the likelihood calculation.}
\end{table}

\begin{figure}[t]
 \centering
 \includegraphics[width=8cm]{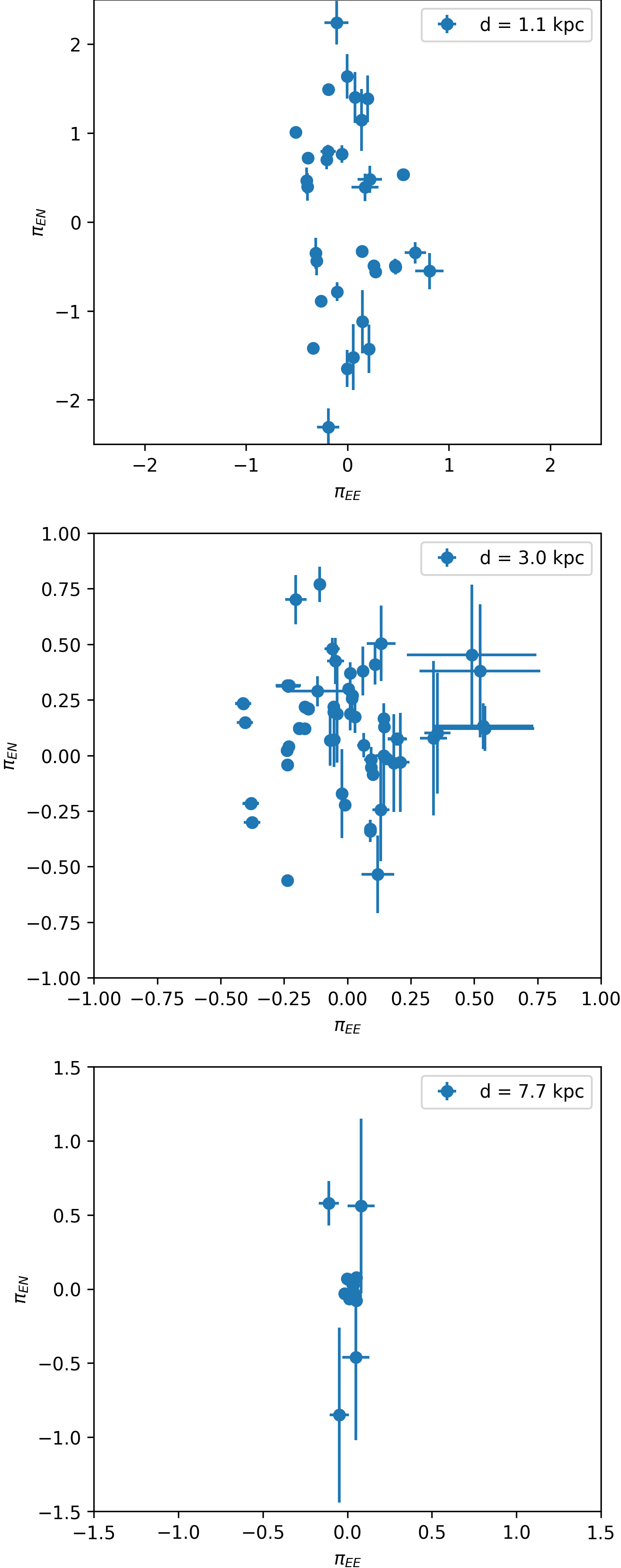} 
 \caption{Distribution of parallax components for microlensing planets in the windows corresponding to the three peaks identified in the anisotropy likelihood.}
 \label{Fig pi_E dist}
 \end{figure}

To this purpose, we have generated plots of the parallax components of all planets contributing to the three peaks in Table \ref{Tab results}, shown in Fig. \ref{Fig pi_E dist}. Indeed, planets in the peak at $d=1.1$ kpc as well as planets in the peak at $d=7.7$ kpc show anomalous distributions, with values of $|\pi_{EN}|$ systematically greater than $|\pi_{EE}|$. It is clear that the ecliptic degeneracy tends to leave large uncertainties in $|\pi_{EN}|$. So, it is possible that by selecting planets with reported measures of parallax we are implicitly favoring large values of $|\pi_{EN}|$, which give measurable effects on the light curves. It is also possible that such behavior is actually due to biases in the procedures adopted for modeling microlensing events. Whatever the reason, it is clear that the peaks at   $d=1.1$ and $d=7.7$ kpc are affected by clear distortions in the sample.

The same Fig. \ref{Fig pi_E dist} shows that the distribution of parallax components for the subsample at $d=3.0$ kpc is free from any evident issues: the two components are generally of similar orders of magnitude as expected from Galactic models. Indeed, the peak at $d=3.0$ kpc comes with a more solid statistics of $n_{eff} = 22.5$ planets, in spite of showing a lower significancy. An anisotropic distribution in planetary position angle with $q=2.8$ is 2.6 times more likely than an isotropic distribution. The fact that the preferred orientation of the angular momenta is $10.3^\circ$ from the Galactic North is suggestive of a real physical effect. Planets 3 kpc away from the Sun seems to have a preferential orientation of their orbital planes close to parallel to the Galactic plane. Of course, such indication would become stronger if we could find an association with a morphological or dynamical structure of our Galaxy. Indeed, a possible interpretation comes from the fact that the the region around $[2,4]$ kpc is dominated by the Scutum-Centaurus arm \citep{Castro2021}, which is one of the major arms in the spiral structure of our Galaxy \citep{Churchwell2009}. Planets inside this overdense structure might be generated with a preferential alignment in their orbital planes with the Galactic plane, while planets in minor structures, such as the Sagittarius and Norma arms, might have a less clear or absent alignment. The consequences of this finding would be very interesting for studies of star and planetary formation in spiral arms: the orientation of orbital planes depends on the environment in which planets form and is affected by the large-scale kinematics of the Galaxy.

\section{Conclusion}

The new method proposed in this paper opens a new avenue for the exploration of anisotropies in our Galaxy exploiting microlensing events. We have shown that using the information retrieved by microlensing modeling it is possible to study the position angles of individual microlensing planets in the sky. The existing sample of planets for which a measure of the proper motion direction has been reported is large enough to start some statistical studies. Indeed we have found that the overall distribution of position angles is fairly isotropic. However, after a closer look at planets in distinct regions along the line of sight to the bulge, we find three anisotropy peaks at distances 1.1, 3 and 7.7 kpc. The first and the last come from a relatively low number of planets and may just be a fluctuation. Alternatively, since these planets are characterized by higher values for the component $\pi_{EN}$, they may signal a possible bias in the microlensing detection efficiency for planetary systems very close to the Sun or very close to the source.

On the other hand, the anisotropy peak at 3 kpc can be interpreted as a a preference for planetary systems in the Scutum-Centaurus arm to have orbital planes aligned with the Galactic plane. In our study we have focused on planetary systems, for which we have access to the complete statistics through the existing databases, but the same method can be applied to binary lensing events as well. Of course, the current statistics is relatively exiguous, but it is going to rapidly increase thanks to future ground and space observations. In particular, the {\it Roman} mission will find more than one thousand planets and will be able to study the dependence on planets occurrence and properties on the Galactocentric distance \cite{Penny2019}. With such homogeneous sample, it will be possible to confirm the tendency discovered in this paper and put it on a more solid statistical basis

Note that the requirement of a non-null parallax measurement may bias our sample toward longer events, for which the annual parallax is easier to measure (although we also include some events with satellite parallax and lens resolution by high-contrast imaging). One may argue that the orientations of the planetary  planes of such systems is not representative for the whole population. This is certainly an interesting question that could find a rigorous answer in the framework of a Galactic model that should be used to check how different is the population of events with measurable annual parallax from the whole sample of microlensing events. Lower relative proper motions may tend to select disk-disk events, since bulge sources have larger velocity dispersions. Then, our planets should generally belong to colder thin disk populations.  

Understanding all possible biases in the statistics of position angles from microlensing observations will probably require several dedicated papers accompanied by much more data. The goal of the present study is limited to the presentation of the general methodology for the derivation of the position angle for any individual case and the proposal of a statistical assessment on the anisotropy of the obtained distribution. A full interpretation of the microlensing observations in the light of these considerations is left to future works.

\begin{acknowledgments}
Authors acknowledge financial support from PRIN2022 CUP D53D23002590006.
\end{acknowledgments}

\appendix

\section{Microlensing planets with measured position angle} \label{Sec Appendix sample}

\begin{longrotatetable}
\begin{deluxetable*}{lcccccccccc}
\tabletypesize{\tiny}
\tablecaption{Microlensing exoplanets included in our investigation. We report the name of each planet, its Galactic coordinates $l, b$, the distance $D_L$ from the Solar System, the projected separation of the planet from the host $a$, the parallax components in North and East direction, and the angle $\alpha$ between the lens axis and the source direction. From these parameters, we derive the position angle $PA$ with respect to the North direction and we report the $\Delta\chi^2$ between the best model and the other degenerate models for every planet. Planets with $^{\dagger}$ symbol benefit from satellite parallax.} 
\tablehead{
\colhead{Event} & \colhead{$l$ (rad)} & \colhead{$b$ (rad)} & \colhead{$D_L$ (kpc)} & \colhead{$s$ (au)} & \colhead{$\pi_N$} & \colhead{$\pi_E$} & \colhead{$\alpha$ (deg)} & \colhead{$PA$ (deg)} & \colhead{$\Delta\chi^2$} & \colhead{Reference}} 
\colnumbers
\startdata 
OGLE-2005-BLG-071Lb & $0.32$ & $1.10$ & $3.46 \pm 0.33$ & $3.54 \pm 0.59$ & $0.12 \pm 0.01$  &  $-0.19 \pm 0.02$ & $94.12 \pm 3.44$ & $-87.10 \pm 3.56$ & $0$ & \citep{OGLE-2005-BLG-071L,OGLE-2005-BLG-071L2}\\
 &  &  &  & $3.53 \pm 0.59$ & $0.12 \pm 0.01$ & $-0.19 \pm 0.02$ & $265.90 \pm 3.44$ & $101.72 \pm 3.56$ & $0.15$ \\
 &  &  &  & $2.09 \pm 0.50$ & $0.12 \pm 0.01$ & $-0.17 \pm 0.02$  & $94.08 \pm 3.44$ & $-83.42 \pm 4.06$ & $4.29$ \\
 &  &  &  & $2.09 \pm 0.50$ & $0.12 \pm 0.01$ &  $-0.19 \pm 0.02$ & $265.93 \pm 3.44$ & $101.65 \pm 3.59$ & $4.31$ \\ 
OGLE-2006-BLG-109Lb & $0.77$ & $0.31$ & $1.51 \pm 0.11$ & $1.42 \pm 0.20$ & $-0.33 \pm 0.02$ &  $0.15 \pm 0.02$ & $201.95 \pm 0.04$ & $20.93 \pm 2.69$ & $0$ & \citep{OGLE-2006-BLG-109L} \\
OGLE-2006-BLG-109Lc & $0.77$ & $0.31$ & $1.51 \pm 0.11$ & $2.37 \pm 0.18$ & $-0.33 \pm 0.02$ &  $0.15 \pm 0.02$ & $21.95 \pm 0.04$ & $200.95 \pm 2.69$ & $0$ & \citep{OGLE-2006-BLG-109L} \\
MOA-2007-BLG-192Lb & $1.16$ & $-0.34$ & $2.16 \pm 0.30$ & $1.68 \pm 0.28$ & $0.32 \pm 0.02$ &  $-0.24 \pm 0.05$ & $120.67 \pm 29.08$ & $-106.22 \pm 5.83$ & $0$ & \citep{MOA-2007-BLG-192} \\
 &  &  &  & $1.90 \pm 0.33$ & $0.32 \pm 0.02$  &  $-0.23 \pm 0.05$ & $110.51 \pm 29.08$ & $-95.49 \pm 5.92$ & $0.03$ \\
 &  &  &  & $1.62 \pm 0.27$ & $0.31 \pm 0.02$ &  $-0.23 \pm 0.05$ & $260.54 \pm 29.08$ & $113.84 \pm 5.90$ & $0.51$ \\
 MOA-2008-BLG-379Lb & $0.79$ & $0.29$ & $3.44 \pm 0.53$ & $2.82 \pm 0.46$ & $0.07 \pm 0.03$  &  $0.20 \pm 0.04$ & $64.82 \pm 0.13$ & $70.71 \pm 7.51$ & $0$ & \citep{MOA-2008-BLG-379L}\\ 
 &  &  &  & $2.82 \pm 0.46$ & $0.08 \pm 0.03$ &  $0.20 \pm 0.04$ & $295.19 \pm 0.13$ & $199.92 \pm 7.57$ & $0.06$ \\
 &  &  &  & $2.40 \pm 0.39$ & $0.07 \pm 0.03$ & $0.20 \pm 0.04$ & $295.03 \pm 0.13$ & $200.80 \pm 7.53$ & $1.93$ \\
 & &  &  & $2.40 \pm 0.39$ &  $0.08 \pm 0.03$ & $0.20 \pm 0.04$ & $64.98 \pm 0.13$ & $70.25 \pm 7.55$ & $2.23$ \\ 
MOA-2009-BLG-266Lb & $-2.32$ & $1.10$ & $3.04 \pm 0.33$ & $2.72 \pm 0.32$ & $0.17 \pm 0.05$ &  $0.144 \pm 0.004$ & $130.02 \pm 0.19$ & $-24.17 \pm 8.59$ & $0$ & \cite{MOA-2009-BLG-266Lb}\\
MOA-2009-BLG-387Lb & $0.54$ & $0.80$ & $5.69 \pm 2.19$ & $1.61 \pm 0.64$ & $2.50 \pm 1.00$ &  $-0.31 \pm 0.30$& $293.08 \pm 0.14$ & $129.04 \pm 7.33$ & $0$ & \citep{MOA-2009-BLG-387L}\\
 &  &  & & $1.61 \pm 0.64$ & $1.70 \pm 1.00$ &  $-0.15 \pm 0.50$ & $293.06 \pm 0.14$ & $131.08 \pm 16.98$ & $4.70$ \\
 MOA-2010-BLG-073Lb & $1.43$ & $-0.58$ & $2.80 \pm 0.40$ & $1.21 \pm 0.26$ & $0.37 \pm 0.05$  & $0.01 \pm 0.01$ & $167.36 \pm 0.40$ & $-128.10 \pm 1.56$ & $0$ & \citep{MOA-2010-BLG-073L}\\
MOA-2010-BLG-117Lb & $1.65$ & $-0.69$ & $3.50 \pm 0.40$ & $2.42 \pm 0.41$ & $-0.17 \pm 0.20$ &  $-0.02 \pm 0.01$ & $112.47 \pm 0.63$ & $101.35 \pm 8.88$ & $0$ & \citep{MOA-2010-BLG-117b} \\
 &  &  &  & $2.33 \pm 0.39$ & $0.19 \pm 0.22$ & $-0.04 \pm 0.01$ & $247.47 \pm 0.57$ & $127.02 \pm 13.96$ & $4.21$ \\ 
MOA-2010-BLG-328Lb & $0.84$ & $0.17$ & $0.81 \pm 0.10$ & $0.92 \pm 0.16$ & $1.01 \pm 0.06$  & $-0.51 \pm 0.04$ & $221.37 \pm 0.50$ & $202.02 \pm 2.27$ & $0$ & \citep{MOA-2010-BLG-328L} \\
 &  &  & $1.24 \pm 0.18$ & $1.21 \pm 0.27$ & $0.72 \pm 0.05$  & $-0.39 \pm 0.03$ & $168.74 \pm 0.87$ & $227.36 \pm 2.49$ & $2.56$ \\
 MOA-2010-BLG-477Lb & $0.71$ & $0.46$ & $2.30 \pm 0.60$ & $3.58 \pm 0.98$ & $0.77 \pm 0.08$  & $-0.11 \pm 0.01$ & $328.94 \pm 3.44$ & $91.02 \pm 1.11$ & $0$ & \citep{MOA-2010-BLG-477L}\\
 &  &  &  & $3.57 \pm 0.97$ & $0.27 \pm 0.03$  & $0.020 \pm 0.002$  & $30.76 \pm 3.44$ & $41.56 \pm 0.64$ & $3.68$ \\
MOA-2011-BLG-028Lb & $0.90$ & $0.06$ & $7.38 \pm 0.52$ & $4.21 \pm 0.73$ & $0.07 \pm 0.01$ & $-0.002 \pm 0.033$ & $125.60 \pm 1.32$ & $-64.62 \pm 26.99$ & $0$ & \citep{MOA-2011-BLG-028L} \\
 &  &  &  & $4.23 \pm 0.73$ &$-0.03 \pm 0.03$  & $-0.02 \pm 0.03$ & $234.70 \pm 1.38$ & $38.89 \pm 48.38$ & $0.0001$ \\ 
OGLE-2011-BLG-0251Lb & $1.15$ & $-0.35$ & $2.57 \pm 0.61$ & $2.71 \pm 1.20$ & $-0.34 \pm 0.05$ & $0.09 \pm 0.01$ & $286.28 \pm 0.11$ & $-69.65 \pm 2.61$ & $0$ & \citep{OGLE-2011-BLG-0251L} \\
 &  &  &  & $1.49 \pm 0.64$ & $-0.33 \pm 0.04$ & $0.09 \pm 0.01$ & $286.28 \pm 0.11$ & $-70.08 \pm 2.39$ & $7$ \\ 
OGLE-2011-BLG-0265Lb & $1.20$ & $-0.39$ & $4.38 \pm 0.48$ & $1.91 \pm 0.28$ &  $0.24 \pm 0.06$ & $0.04 \pm 0.02$ & $27.15 \pm 0.14$ & $32.14 \pm 4.68$ & $0$ & \citep{OGLE-2011-BLG-0265L} \\
 &  &  & $3.49 \pm 0.60$ & $1.53 \pm 0.30$ & $0.38 \pm 0.11$ & $0.06 \pm 0.02$ & $334.04 \pm 0.23$ & $84.38 \pm 3.50$ & $5.70$ \\
OGLE-2012-BLG-0406Lb & $0.81$ & $0.23$ & $4.97 \pm 0.29$ & $3.43 \pm 0.38$ & $-0.14 \pm 0.02$ & $0.05 \pm 0.01$ & $312.04 \pm 0.11$ & $-84.88 \pm 3.32$ & $0$ & \citep{OGLE-2012-BLG-0406L,OGLE-2012-BLG-0406L2}\\
OGLE-2012-BLG-0358Lb & $1.94$ & $-0.77$ & $1.76 \pm 0.13$ & $0.87 \pm 0.11$ & $1.49 \pm 0.07$ & $-0.19 \pm 0.06$ & $41.37 \pm 0.06$ & $-37.48 \pm 2.30$ & $0$ & \citep{OGLE-2012-BLG-0358L} \\
 &  &  & $1.79 \pm 0.12$ & $0.88 \pm 0.11$ & $-1.42 \pm 0.06$ & $-0.34 \pm 0.04$ & $318.69 \pm 0.06$ & $-114.06 \pm 1.62$ & $1.98$ \\ 
OGLE-2012-BLG-0026Lb & $1.15$ & $-0.35$ & $4.02 \pm 0.38$ & $4.07 \pm 0.42$ & $-0.07 \pm 0.05$  & $0.114 \pm 0.004$ & $253.58 \pm 0.06$ & $-79.87 \pm 18.35$ & $0$ & \citep{OGLE-2012-BLG-0026L,OGLE-2012-BLG-0026Lb}\\
 &  &  &  & $3.77 \pm 0.39$ & $0.001 \pm 0.028$ &  $0.12 \pm 0.01$  & $106.43 \pm 0.06$ & $34.53 \pm 13.04$ & $6.40$ \\
 &  &  &  & $4.07 \pm 0.42$ & $-0.10 \pm 0.01$  & $0.137 \pm 0.003$  & $106.43 \pm 0.06$ & $70.29 \pm 3.40$ & $6.60$ \\ 
OGLE-2012-BLG-0026Lc & $1.15$ & $-0.35$ & $4.02 \pm 0.38$ & $4.94 \pm 0.51$ & $-0.07 \pm 0.05$ & $0.114 \pm 0.004$  & $30.64 \pm 0.06$ & $143.07 \pm 18.35$ & $0$ & \citep{OGLE-2012-BLG-0026L,OGLE-2012-BLG-0026Lb} \\
 &  &  &  & $4.95 \pm 0.51$ & $0.001 \pm 0.028$  & $0.12 \pm 0.01$ & $329.44 \pm 0.06$ & $171.53 \pm 13.04$ & $6.40$ \\ 
 &  &  &  & $3.23 \pm 0.33$ & $-0.10 \pm 0.01$  & $0.137 \pm 0.003$ & $329.67 \pm 0.06$ & $207.07 \pm 3.40$ & $6.60$ \\ 
OGLE-2012-BLG-0950Lb & $0.96$ & $-0.05$ & $2.19 \pm 0.23$ & $2.48 \pm 0.34$ & $0.21 \pm 0.02$  & $-0.15 \pm 0.02$ & $111.49 \pm 0.34$ & $-87.38 \pm 3.60$ & $0$ & \citep{OGLE-2012-BLG-0950L} \\
 &  &  &  & $2.76 \pm 0.37$ & $0.21 \pm 0.02$  & $-0.15 \pm 0.02$ & $111.57 \pm 0.34$ & $-87.49 \pm 3.60$ & $1.62$ \\
  &  &  &  & $2.48 \pm 0.34$ & $0.22 \pm 0.02$  & $-0.17 \pm 0.02$  & $248.85 \pm 0.34$ & $133.83 \pm 3.40$ & $1.63$ \\ 
 &  &  &  & $2.76 \pm 0.37$ & $0.22 \pm 0.02$  & $-0.17 \pm 0.02$ & $248.76 \pm 0.34$ & $133.92 \pm 3.42$ & $3.20$ \\ 
OGLE-2013-BLG-0102Lb & $0.73$ & $0.41$ & $3.04 \pm 0.31$ & $0.79 \pm 0.11$ & $0.48 \pm 0.05$  & $0.06 \pm 0.03$ & $184.42 \pm 1.15$ & $-109.83 \pm 3.60$ & $0$ & \citep{OGLE-2013-BLG-0102L} \\
OGLE-2013-BLG-0132b & $1.05$ & $-0.20$ & $3.48 \pm 0.36$ & $3.14 \pm 0.33$ & $0.13 \pm 0.02$  & $0.15 \pm 0.02$ & $132.98 \pm 0.34$ & $-28.27 \pm 6.16$ & $0$ & \citep{OGLE-2013-BLG-0132} \\
OGLE-2013-BLG-0911Lb &  &  &  & $0.32 \pm 0.06$ &  $0.26 \pm 0.05$  & $0.02 \pm 0.01$ & $290.53 \pm 0.23$ & $126.91 \pm 1.36$ & $0$ & \citep{OGLE-2013-BLG-0911L}\\
 &  &  & & $0.32 \pm 0.06$ & $0.30 \pm 0.03$  & $0.004 \pm 0.006$ & $119.06 \pm 0.40$ & $-55.87 \pm 1.15$ & $1.10$ \\ 
MOA-2013-BLG-605Lb & $0.92$ & $0.02$ & $0.85 \pm 0.08$ & $0.92 \pm 0.11$ & $-2.31 \pm 0.21$  & $-0.19 \pm 0.11$ & $349.88 \pm 0.58$ & $-103.40 \pm 2.70$ & $0$ & \citep{MOA-2013-BLG-605L} \\
 &  &  & $0.86 \pm 0.09$ & $0.96 \pm 0.12$ & $2.24 \pm 0.25$  & $-0.11 \pm 0.12$ & $2.72 \pm 0.59$ & $56.29 \pm 3.00$ & $2.59$ \\
  &  &  & $1.92 \pm 0.20$ & $2.22 \pm 0.28$ & $0.79 \pm 0.07$  & $-0.19 \pm 0.08$ & $359.43 \pm 0.34$ & $48.84 \pm 5.33$ & $5.83$ \\ 
 &  &  & $3.55 \pm 0.49$ & $4.19 \pm 0.64$ & $0.29 \pm 0.07$ & $-0.12 \pm 0.10$  & $359.13 \pm 0.33$ & $40.46 \pm 18.13$ & $7.64$ \\ 
 &  &  & $1.76 \pm 0.17$ & $2.01 \pm 0.24$ & $-0.89 \pm 0.07$  & $-0.26 \pm 0.06$ & $0.20 \pm 0.15$ & $-102.22 \pm 3.70$ & $8.10$ \\
 OGLE-2014-BLG-0124Lb$^{\dagger}$ & $1.05$ & $-0.19$ & $3.50 \pm 0.20$ & $3.40 \pm 0.28$ & $-0.01 \pm 0.01$  & $0.145 \pm 0.004$ & $281.60 \pm 0.11$ & $226.88 \pm 1.98$ & $0$ & \citep{OGLE-2014-BLG-0124Lb,OGLE-2014-BLG-0124Lb2} \\
 &  &  &  & $3.40 \pm 0.28$ & $-0.015 \pm 0.002$  & $0.16 \pm 0.01$ & $78.30 \pm 0.11$ & $73.72 \pm 0.77$ & $3.50$ \\ 
OGLE-2015-BLG-0966Lb$^{\dagger}$ & $0.88$ & $0.10$ & $2.90 \pm 0.40$ & $2.46 \pm 0.41$ & $-0.041 \pm 0.001$  & $-0.24 \pm 0.01$ & $230.40 \pm 0.10$ & $93.03 \pm 0.40$ & $0$ & \citep{OGLE-2015-BLG-0966L} \\
 &  &  &  & $2.00 \pm 0.33$ & $-0.041 \pm 0.001$  & $-0.24 \pm 0.01$ & $230.40 \pm 0.10$ & $93.03 \pm 0.35$ & $0.0001$ \\ 
 & &  &  & $2.00 \pm 0.33$ & $0.023 \pm 0.001$  & $-0.24 \pm 0.01$ & $129.40 \pm 0.10$ & $209.50 \pm 0.29$ & $1$ \\ 
 &  &  &  & $2.46 \pm 0.41$ &  $0.023 \pm 0.001$  & $-0.24 \pm 0.01$ & $129.30 \pm 0.10$ & $209.58 \pm 0.33$ & $1$ \\ 
 &  &  &  & $2.46 \pm 0.41$ &  $0.040 \pm 0.001$  & $-0.232 \pm 0.003$ & $230.40 \pm 0.10$ & $112.60 \pm 0.33$ & $2$ \\ 
 &  &  &  & $2.46 \pm 0.41$ & $-0.561 \pm 0.002$  & $-0.24 \pm 0.01$  & $129.40 \pm 0.10$ & $136.81 \pm 1.05$ & $3$ \\ 
K2-2016-BLG-005Lb$^{\dagger}$ & $1.24$ & $-0.43$ & $5.20 \pm 0.24$ & $4.19 \pm 0.29$ & $-0.110 \pm 0.003$  & $-0.045 \pm 0.002$ & $121.86 \pm 4.60$ & $127.80 \pm 0.94$ & $0$ & \citep{K2-2016-BLG-005}\\
KMT-2016-BLG-1836Lb & $0.77$ & $0.33$ & $7.20 \pm 1.50$ & $1.11 \pm 4.59$ & $-0.46 \pm 0.56$  & $0.05 \pm 0.08$ & $205.31 \pm 1.70$ & $35.13 \pm 12.38$ & $0$ & \citep{KMT-2016-BLG-1836L}\\
 &  &  &  & $1.11 \pm 4.56$ &  $0.56 \pm 0.59$  & $0.08 \pm 0.08$ & $155.10 \pm 2.01$ & $-80.35 \pm 11.65$ & $0.50$ \\ 
OGLE-2016-BLG-1093Lb$^{\dagger}$ & $0.63$ & $0.63$ & $8.12 \pm 0.69$ & $1.98 \pm 0.37$ & $-0.044 \pm 0.003$  & $0.05 \pm 0.01$ & $325.10 \pm 1.49$ & $237.10 \pm 4.14$ & $0$ & \citep{OGLE-2016-BLG-1093L}\\
 &  &  & $8.06 \pm 0.91$ & $2.05 \pm 0.40$ & $0.058 \pm 0.002$  & $0.03 \pm 0.01$ & $34.78 \pm 1.43$ & $57.62 \pm 5.05$ & $2.12$ \\ 
OGLE-2016-BLG-1190Lb$^{\dagger}$ & $1.24$ & $-0.43$ & $6.77 \pm 0.08$ & $2.00 \pm 0.17$ & $0.067 \pm 0.003$  & $0.004 \pm 0.006$ & $182.18 \pm 0.29$ & $-131.42 \pm 5.12$ & $0$ & \citep{OGLE-2016-BLG-1190L}\\
 &  &  & $6.74 \pm 0.08$ & $1.99 \pm 0.16$ & $-0.066 \pm 0.002$  & $0.01 \pm 0.01$ & $177.77 \pm 0.29$ & $39.27 \pm 5.05$ & $3.78$ \\ 
OGLE-2016-BLG-1067Lb$^{\dagger}$& $1.14$ & $-0.31$ & $3.68 \pm 0.65$ & $1.54 \pm 0.72$ & $0.22 \pm 0.02$  & $-0.05 \pm 0.01$ & $98.24 \pm 0.34$ & $-59.79 \pm 2.17$ & $0$ & \citep{OGLE-2016-BLG-1067L} \\
 &  &  & $3.78 \pm 0.69$ & $1.52 \pm 0.78$ & $0.20 \pm 0.02$  & $-0.05 \pm 0.01$ & $98.07 \pm 0.34$ & $-61.23 \pm 2.72$ & $5$ \\ 
 &  &  & $4.30 \pm 0.65$ & $1.14 \pm 0.78$ & $-0.22 \pm 0.01$  & $-0.01 \pm 0.01$ & $261.80 \pm 0.34$ & $-26.95 \pm 1.93$ & $6$ \\ 
OGLE-2016-BLG-1195Lb$^{\dagger}$ & $0.79$ & $0.29$ & $3.91 \pm 0.44$ & $1.10 \pm 0.20$ & $-0.30 \pm 0.01$  & $-0.38 \pm 0.03$ & $124.52 \pm 0.13$ & $172.92 \pm 2.47$ & $0$ & \citep{OGLE-2016-BLG-1195L}\\
 &  &  &  & $1.10 \pm 0.20$ &  $0.15 \pm 0.01$ &  $-0.40 \pm 0.03$ & $124.52 \pm 0.13$ & $231.87 \pm 1.67$ & $0.0001$ \\ 
 &  &  &  & $1.21 \pm 0.22$ & $-0.30 \pm 0.01$  & $-0.38 \pm 0.03$ & $124.49 \pm 0.13$ & $173.00 \pm 2.46$ & $0.0001$ \\ 
 &  &  &  & $1.10 \pm 0.20$ & $0.23 \pm 0.01$  & $-0.41 \pm 0.03$ & $235.47 \pm 0.13$ & $130.28 \pm 2.10$ & $1$ \\ 
 &  &  &  & $1.10 \pm 0.20$ & $-0.22 \pm 0.01$ & $-0.38 \pm 0.03$  & $235.49 \pm 0.13$ & $71.13 \pm 2.22$ & $1$ \\ 
 &  &  &  & $1.21 \pm 0.22$ & $0.15 \pm 0.01$  & $-0.40 \pm 0.03$ & $124.49 \pm 0.13$ & $231.92 \pm 1.78$ & $1$ \\ 
 &  &  &  & $1.21 \pm 0.22$ & $0.24 \pm 0.01$  & $-0.41 \pm 0.03$ & $235.51 \pm 0.13$ & $130.32 \pm 2.03$ & $1$ \\ 
 &  &  &  & $1.21 \pm 0.22$ & $-0.22 \pm 0.01$  & $-0.38 \pm 0.03$ & $235.52 \pm 0.13$ & $71.16 \pm 2.21$ & $1$ \\ 
KMT-2017-BLG-0165Lb & $1.01$ & $-0.14$ & $4.53 \pm 1.01$ & $3.45 \pm 0.95$ & $-0.10 \pm 0.47$  & $0.11 \pm 0.04$ & $206.48 \pm 1.89$ & $-15.91 \pm 136.19$ & $0$ & \citep{KMT-2017-BLG-0165L}\\
 &  &  &  & $3.45 \pm 0.95$ & $0.05 \pm 0.46$  & $0.09 \pm 0.04$ & $153.29 \pm 1.83$ & $-34.30 \pm 227.41$ & $0.50$ \\
 OGLE-2017-BLG-0406$b^{\dagger}$ & $0.97$ & $-0.08$ & $5.20 \pm 0.50$ & $3.48 \pm 0.34$ & $0.12 \pm 0.02$  & $0.07 \pm 0.01$ & $233.46 \pm 0.06$ & $214.51 \pm 5.45$ & $0$ & \citep{OGLE-2017-BLG-0406} \\
 &  &  &  & $3.48 \pm 0.34$ & $0.13 \pm 0.02$  & $0.06 \pm 0.01$ & $123.11 \pm 0.06$ & $-37.38 \pm 4.57$ & $2.62$ \\ 
OGLE-2017-BLG-1140Lb$^{\dagger}$ & $2.03$ & $-0.78$ & $7.36 \pm 0.01$ & $1.04 \pm 0.14$ & $0.079 \pm 0.001$  & $0.053 \pm 0.001$ & $34.54 \pm 0.39$ & $5.54 \pm 0.62$ & $0$ & \citep{OGLE-2017-BLG-1140L} \\
 &  &  &  & $1.05 \pm 0.15$ & $-0.078 \pm 0.002$  & $0.053 \pm 0.001$  & $34.53 \pm 0.43$ & $117.40 \pm 0.68$ & $1$ \\
 OGLE-2017-BLG-1434Lb & $0.79$ & $0.28$ & $0.86 \pm 0.05$ & $1.18 \pm 0.10$ & $-0.49 \pm 0.08$  & $0.47 \pm 0.01$ & $279.19 \pm 0.06$ & $-76.97 \pm 4.67$ & $0$ & \citep{OGLE-2017-BLG-1434L}\\
 &  &  &  & $1.18 \pm 0.10$ & $-0.51 \pm 0.08$  & $0.48 \pm 0.01$ & $81.04 \pm 0.11$ & $121.95 \pm 4.74$ & $3.77$ \\ 
OGLE-2017-BLG-1275Lb$^{\dagger}$ & $0.93$ & $-0.00$ & $7.72 \pm 0.82$ & $0.32 \pm 0.25$ & $-0.03 \pm 0.02$  & $0.03 \pm 0.01$ & $267.58 \pm 0.75$ & $-71.23 \pm 19.03$ & $0$ & \citep{OGLE-2017-BLG-1275}\\
 &  &  & $7.65 \pm 0.90$ & $0.32 \pm 0.24$ & $0.03 \pm 0.02$  & $0.03 \pm 0.02$ & $267.64 \pm 0.75$ & $194.18 \pm 19.78$ & $1.15$ \\ 
 &  &  & $7.69 \pm 0.89$ & $0.22 \pm 0.16$ & $0.03 \pm 0.02$  & $0.03 \pm 0.02$ & $267.47 \pm 0.75$ & $205.43 \pm 20.86$ & $3.09$ \\ 
 &  &  & $7.74 \pm 0.82$ & $0.25 \pm 0.19$ & $-0.03 \pm 0.02$  & $0.04 \pm 0.02$  & $267.47 \pm 0.80$ & $-81.34 \pm 20.50$ & $3.32$ \\ 
OGLE-2017-BLG-1375Lb & $0.80$ & $0.26$ & $3.93 \pm 1.20$ & $3.26 \pm 1.09$ & $0.07 \pm 0.12$  & $-0.05 \pm 0.02$ & $292.32 \pm 0.40$ & $97.70 \pm 47.40$ & $0$ & \citep{OGLE-2017-BLG-1375L} \\
 &  &  & $3.79 \pm 1.02$ & $3.15 \pm 0.94$ & $0.07 \pm 0.11$  & $-0.07 \pm 0.03$ & $67.68 \pm 0.52$ & $-46.39 \pm 48.48$ & $0.10$ \\ 
 &  &  & $4.07 \pm 1.10$ & $5.11 \pm 1.54$ & $0.13 \pm 0.11$  & $0.004 \pm 0.121$ & $69.63 \pm 0.34$ & $-2.04 \pm 51.71$ & $7.60$ \\ 
 &  &  & $4.08 \pm 1.14$ & $5.12 \pm 1.58$ & $-0.06 \pm 0.02$  & $-0.04 \pm 0.12$ & $290.43 \pm 0.29$ & $-11.30 \pm 76.36$ & $7.80$ \\ 
OGLE-2017-BLG-1806Lb & $1.94$ & $-0.77$ & $6.17 \pm 1.25$ & $1.66 \pm 0.68$ & $0.29 \pm 0.17$  & $0.14 \pm 0.06$ & $179.79 \pm 2.12$ & $-142.61 \pm 16.09$ & $0$ & \citep{KMT-2021-BLG-0119}\\
 &  &  & $6.60 \pm 0.86$ & $1.83 \pm 0.69$ & $-0.28 \pm 0.15$  & $0.11 \pm 0.06$ & $180.15 \pm 1.95$ & $-9.89 \pm 14.27$ & $0.20$ \\ 
 &  &  & $2.87 \pm 0.99$ & $5.77 \pm 4.05$ & $0.50 \pm 0.17$  & $0.13 \pm 0.06$ & $358.82 \pm 2.06$ & $26.87 \pm 7.63$ & $8.30$ \\ 
 &  &  & $3.01 \pm 1.66$ & $5.44 \pm 5.15$ & $-0.54 \pm 0.18$  & $0.12 \pm 0.07$ & $1.09 \pm 1.95$ & $177.20 \pm 7.74$ & $8.40$ \\ 
KMT-2018-BLG-0029Lb$^{\dagger}$ & $1.31$ & $-0.50$ & $3.21 \pm 0.25$ & $5.01 \pm 0.86$ & $-0.09 \pm 0.03$  & $0.10 \pm 0.01$ & $267.72 \pm 0.29$ & $-93.13 \pm 9.93$ & $0$ & \citep{KMT-2018-BLG-0029L}\\
 &  &  & $3.52 \pm 0.24$ & $5.49 \pm 0.92$ & $-0.05 \pm 0.04$  & $0.09 \pm 0.02$ & $92.11 \pm 0.29$ & $71.93 \pm 19.82$ & $4.30$ \\
KMT-2018-BLG-1292Lb & $0.47$ & $0.91$ & $3.46 \pm 0.47$ & $8.11 \pm 1.82$ &  $-0.02 \pm 0.06$  & $0.09 \pm 0.03$ & $329.14 \pm 0.97$ & $199.96 \pm 33.99$ & $0$ & \citep{KMT-2018-BLG-1292L}\\
 &  &  & $3.84 \pm 0.61$ & $8.93 \pm 2.14$ & $0.05 \pm 0.06$  & $0.06 \pm 0.03$ & $31.43 \pm 0.97$ & $91.01 \pm 34.29$ & $0.68$ \\ 
KMT-2018-BLG-1988Lb &  &  & $4.15 \pm 1.60$ & $9.93 \pm 14.58$ &$-0.57 \pm 0.76$  & $0.19 \pm 0.10$ & $243.35 \pm 1.20$ & $-36.49 \pm 24.64$ & $0$ & \citep{KMT-2018-BLG-1988L} \\
 &  &  & $4.17 \pm 1.62$ & $4.57 \pm 12.05$ &  $-0.23 \pm 0.75$  & $0.19 \pm 0.10$ & $243.35 \pm 1.32$ & $-57.61 \pm 92.92$ & $0.60$ \\ 
OGLE-2018-BLG-0532Lb & $1.14$ & $-0.31$ & $0.73 \pm 0.10$ & $1.00 \pm 0.15$ & $-0.79 \pm 0.11$  & $-0.10 \pm 0.02$ & $206.64 \pm 0.29$ & $33.32 \pm 1.63$ & $0$ & \citep{OGLE-2018-BLG-0532L} \\
 &  &  & $0.80 \pm 0.10$ & $1.09 \pm 0.15$ & $0.77 \pm 0.10$  & $-0.05 \pm 0.02$ & $153.36 \pm 0.29$ & $-104.98 \pm 1.23$ & $0.58$ \\ 
OGLE-2018-BLG-0596Lb$^{\dagger}$ & $0.89$ & $0.06$ & $5.65 \pm 0.75$ & $0.97 \pm 0.17$ & $-0.02 \pm 0.02$  & $0.18 \pm 0.01$ & $194.97 \pm 0.63$ & $-34.96 \pm 6.98$ & $0$ & \citep{OGLE-2018-BLG-0596} \\
 &  &  &  & $1.03 \pm 0.18$ & $0.04 \pm 0.02$  & $0.18 \pm 0.01$ & $167.93 \pm 0.63$ & $-28.78 \pm 6.70$ & $5$ \\ 
OGLE-2018-BLG-0799Lb$^{\dagger}$ & $1.69$ & $-0.71$ & $3.93 \pm 2.00$ & $0.61 \pm 0.93$ & $0.41 \pm 0.09$  & $0.11 \pm 0.02$ & $247.00 \pm 0.17$ & $152.15 \pm 4.33$ & $0$ & \citep{OGLE-2018-BLG-0799L}\\
 &  &  & $6.19 \pm 0.45$ & $0.96 \pm 1.38$ & $-0.30 \pm 0.08$  & $0.15 \pm 0.01$ & $113.16 \pm 0.17$ & $64.31 \pm 6.17$ & $3.40$ \\ 
OGLE-2018-BLG-1212Lb & $1.02$ & $-0.15$ & $1.55 \pm 0.91$ & $2.24 \pm 2.03$ & $0.53 \pm 0.02$ &  $0.55 \pm 0.01$ & $296.69 \pm 0.34$ & $166.83 \pm 1.17$ & $0$ & \citep{OGLE-2018-BLG-1212} \\
 &  &  &  & $1.05 \pm 0.95$ & $0.53 \pm 0.02$  & $0.55 \pm 0.01$ & $296.75 \pm 0.34$ & $166.83 \pm 1.17$ & $3.37$ \\ 
OGLE-2018-BLG-1269Lb & $1.24$ & $-0.43$ & $2.51 \pm 0.76$ & $4.52 \pm 1.41$ &  $0.19 \pm 0.07$  &  $0.01 \pm 0.02$ & $71.88 \pm 3.90$ & $-21.22 \pm 4.77$ & $0$ & \citep{OGLE-2018-BLG-1269L}\\
 &  &  & $2.64 \pm 0.79$ & $4.75 \pm 1.47$ & $0.17 \pm 0.07$  & $0.03 \pm 0.01$ & $288.35 \pm 3.95$ & $128.42 \pm 5.44$ & $0.90$ \\ 
KMT-2018-BLG-1990Lb & $2.58$ & $-0.72$ & $0.97 \pm 0.26$ & $0.82 \pm 0.24$ & $1.39 \pm 0.26$  & $0.20 \pm 0.05$ & $322.53 \pm 1.20$ & $68.78 \pm 2.50$ & $0$ & \citep{KMT-2018-BLG-1990L} \\
 &  &  & $0.87 \pm 0.21$ & $0.74 \pm 0.20$ & $-1.43 \pm 0.27$  & $0.21 \pm 0.05$ & $37.36 \pm 0.97$ & $157.40 \pm 2.40$ & $0.06$ \\ 
 &  &  & $0.74 \pm 0.17$ & $0.71 \pm 0.19$ & $1.64 \pm 0.25$  & $-0.002 \pm 0.043$ & $323.27 \pm 0.92$ & $59.83 \pm 1.51$ & $0.48$ \\ 
 &  &  & $1.74 \pm 0.77$ & $1.47 \pm 0.68$ & $1.15 \pm 0.35$  & $0.14 \pm 0.03$ & $31.95 \pm 1.03$ & $-1.85 \pm 2.60$ & $0.89$ \\ 
 &  &  & $0.67 \pm 0.13$ & $0.65 \pm 0.15$ & $-1.65 \pm 0.21$  & $-0.003 \pm 0.038$  & $36.44 \pm 0.63$ & $166.83 \pm 1.32$ & $1.07$ \\ 
 &  &  & $1.28 \pm 1.03$ & $1.08 \pm 0.88$ & $-1.12 \pm 0.36$ & $0.15 \pm 0.03$ & $327.97 \pm 1.15$ & $-132.21 \pm 2.88$ & $1.71$ \\ 
 &  &  & $0.94 \pm 0.35$ & $0.92 \pm 0.36$ & $-1.52 \pm 0.37$  & $0.06 \pm 0.03$ & $325.05 \pm 0.86$ & $-123.98 \pm 1.38$ & $3.77$ \\ 
 &  &  & $0.95 \pm 0.32$ & $0.93 \pm 0.34$ & $1.40 \pm 0.29$  & $0.08 \pm 0.05$  & $34.95 \pm 0.86$ & $-8.72 \pm 2.02$ & $4.61$ \\ 
KMT-2019-BLG-0298Lb & $1.24$ & $-0.44$ & $6.56 \pm 1.54$ & $37.71 \pm 11.43$ & $0.54 \pm 0.44$  & $0.15 \pm 0.13$ & $48.97 \pm 3.27$ & $14.07 \pm 17.64$ & $0$ & \citep{KMT-2019-BLG-0298} \\
 &  &  & $6.92 \pm 1.19$ & $63.27 \pm 14.14$ & $-0.85 \pm 0.59$ &  $-0.05 \pm 0.06$ & $343.49 \pm 3.95$ & $-112.94 \pm 4.46$ & $0.10$ \\ 
KMT-2019-BLG-1552Lb & $1.23$ & $-0.42$ & $4.60 \pm 1.55$ & $3.17 \pm 3.20$ & $0.11 \pm 0.15$  & $0.08 \pm 0.01$ & $281.96 \pm 0.74$ & $162.54 \pm 37.69$ & $0$ & \citep{KMT-2019-BLG-1552} \\
 &  &  &  & $2.16 \pm 4.47$ & $0.02 \pm 0.19$  & $0.09 \pm 0.01$ & $77.50 \pm 1.03$ & $48.06 \pm 111.22$ & $1.20$ \\
KMT-2019-BLG-1216Lb & $-0.58$ & $1.40$ & $2.90 \pm 0.99$ & $5.87 \pm 2.22$ & $0.12 \pm 0.10$  & $0.54 \pm 0.19$ & $91.88 \pm 1.26$ & $7.41 \pm 11.05$ & $0$ & \citep{KMT-2019-BLG-0298} \\
 &  &  & $2.93 \pm 1.00$ & $5.76 \pm 2.17$ & $0.13 \pm 0.10$  & $0.54 \pm 0.20$ & $91.85 \pm 1.26$ & $6.27 \pm 11.52$ & $0.0001$ \\ 
 &  &  & $2.61 \pm 1.02$ & $5.38 \pm 2.44$ & $0.38 \pm 0.30$  & $0.52 \pm 0.24$ & $88.33 \pm 1.15$ & $-12.44 \pm 24.74$ & $0.10$ \\ 
 &  &  & $2.65 \pm 1.06$ & $5.58 \pm 2.57$ & $0.45 \pm 0.32$  & $0.49 \pm 0.26$ & $88.48 \pm 1.15$ & $-19.24 \pm 24.86$ & $0.20$ \\ 
KMT-2019-BLG-1806Lb & $0.92$ & $0.02$ & $6.62 \pm 1.42$ & $2.63 \pm 6.80$ &  $-0.02 \pm 0.02$  & $-0.06 \pm 0.01$& $303.16 \pm 0.46$ & $11.00 \pm 15.18$ & $0$ & \citep{KMT-2021-BLG-0119} \\
 &  &  & $6.68 \pm 1.35$ & $3.40 \pm 6.59$ & $-0.06 \pm 0.16$  & $-0.06 \pm 0.16$ & $303.33 \pm 0.52$ & $-18.07 \pm 109.47$ & $0.40$ \\ 
 &  &  & $6.62 \pm 1.33$ & $3.03 \pm 5.84$ & $-0.06 \pm 0.15$  & $-0.06 \pm 0.02$ & $303.19 \pm 0.46$ & $-14.97 \pm 78.47$ & $0.70$ \\ 
 &  &  & $6.63 \pm 1.37$ & $3.32 \pm 6.20$ & $-0.07 \pm 0.16$  & $-0.06 \pm 0.01$ & $300.63 \pm 0.46$ & $-17.11 \pm 69.77$ & $1.10$ \\ 
OGLE-2019-BLG-0249Lb & $0.32$ & $1.10$ & $6.33 \pm 0.85$ & $1.61 \pm 0.73$ & $0.02 \pm 0.08$  & $0.06 \pm 0.02$ & $215.73 \pm 0.69$ & $-75.66 \pm 66.93$ & $0$ & \citep{KMT-2019-BLG-0298} \\
 &  &  & $6.36 \pm 0.83$ & $1.53 \pm 0.74$ & $0.01 \pm 0.08$  & $0.06 \pm 0.02$ & $241.70 \pm 0.69$ & $-94.64 \pm 78.24$ & $0.10$ \\ 
 &  &  & $6.63 \pm 0.66$ & $4.66 \pm 2.46$ & $-0.003 \pm 0.068$ &  $0.06 \pm 0.02$ & $236.31 \pm 0.23$ & $-78.47 \pm 69.38$ & $3.20$ \\
  &  &  & $6.69 \pm 0.66$ & $4.62 \pm 2.48$ & $-0.01 \pm 0.08$  & $0.06 \pm 0.02$ & $236.13 \pm 0.29$ & $35.98 \pm 76.15$ & $3.20$ \\
 OGLE-2019-BLG-0960Lb & $1.78$ & $-0.73$ & $0.98 \pm 0.21$ & $1.92 \pm 0.43$ & $-0.35 \pm 0.17$  & $-0.31 \pm 0.03$ & $164.42 \pm 0.06$ & $76.86 \pm 14.24$ & $0$ & \citep{OGLE-2019-BLG-0960L} \\
 &  &  & $0.81 \pm 0.17$ & $1.67 \pm 0.36$ & $-0.44 \pm 0.16$  &$-0.30 \pm 0.03$  & $195.44 \pm 0.06$ & $38.77 \pm 10.09$ & $1$ \\
  &  &  & $0.70 \pm 0.13$ & $1.46 \pm 0.28$ & $0.46 \pm 0.15$  & $-0.41 \pm 0.04$ & $164.32 \pm 0.06$ & $174.13 \pm 9.34$ & $1.90$ \\
  &  &  & $0.83 \pm 0.17$ & $1.65 \pm 0.36$ & $0.40 \pm 0.16$  & $-0.39 \pm 0.04$ & $164.36 \pm 0.06$ & $170.33 \pm 11.60$ & $2.10$ \\ 
KMT-2020-BLG-0414Lb & $1.06$ & $-0.21$ & $0.80 \pm 0.11$ & $1.27 \pm 0.18$ & $-0.35 \pm 0.12$  & $0.67 \pm 0.11$ & $358.46 \pm 0.03$ & $174.78 \pm 8.91$ & $0$ & \citep{KMT-2020-BLG-0414L}\\
 &  &  & $0.69 \pm 0.12$ & $1.05 \pm 0.19$ & $-0.55 \pm 0.20$  & $0.81 \pm 0.14$ & $358.66 \pm 0.03$ & $181.58 \pm 10.84$ & $1.50$ \\
 &  &  & $1.22 \pm 0.27$ & $1.80 \pm 0.41$ & $0.39 \pm 0.15$  & $0.17 \pm 0.14$ & $1.46 \pm 0.06$ & $78.46 \pm 18.62$ & $1.70$ \\ 
 &  &  & $1.12 \pm 0.23$ & $1.62 \pm 0.34$ &  $0.48 \pm 0.15$ & $0.22 \pm 0.12$ & $1.27 \pm 0.03$ & $79.30 \pm 13.77$ & $6.80$ \\ 
KMT-2021-BLG-0322Lb & $0.94$ & $-0.02$ & $6.60 \pm 1.00$ & $3.26 \pm 0.56$ & $-2.31 \pm 2.05$ & $-0.28 \pm 0.18$  & $86.10 \pm 0.46$ & $161.63 \pm 7.50$ & $0$ & \citep{KMT-2021-BLG-0322} \\
KMT-2021-BLG-0712Lb & $0.68$ & $0.52$ & $2.08 \pm 0.54$ & $1.49 \pm 0.45$ &  $0.70 \pm 0.11$  & $-0.20 \pm 0.04$ & $155.00 \pm 2.24$ & $-102.71 \pm 3.99$ & $0$ & \citep{KMT-2021-BLG-0712} \\
 &  &  & $3.20 \pm 0.80$ & $2.43 \pm 0.70$ & $0.43 \pm 0.10$ & $-0.05 \pm 0.03$  & $154.10 \pm 1.75$ & $-91.93 \pm 4.78$ & $1.97$ \\ 
KMT-2021-BLG-0912Lb & $0.76$ & $0.34$ & $5.56 \pm 1.43$ & $3.56 \pm 1.29$ & $0.15 \pm 0.28$  & $0.00\pm 0.09$ & $354.04 \pm 0.46$ & $72.74 \pm 34.38$ & $0$ & \citep{KMT-2021-BLG-0912L} \\
 &  &  & $6.71 \pm 1.11$ & $2.90 \pm 0.72$ & $0.58 \pm 0.15$  & $-0.11 \pm 0.06$ & $355.24 \pm 0.69$ & $60.80 \pm 6.33$ & $9.50$ \\ 
KMT-2021-BLG-2478Lb & $0.88$ & $0.09$ & $3.06 \pm 0.71$ & $1.73 \pm 0.42$ & $0.08 \pm 0.35$  & $0.34 \pm 0.05$ & $164.92 \pm 0.57$ & $-24.79 \pm 55.89$ & $0$ & \citep{KMT-2021-BLG-0712} \\
 &  &  & $2.86 \pm 0.68$ & $1.75 \pm 0.43$ & $0.10 \pm 0.27$  &$0.36 \pm 0.05$ & $160.73 \pm 0.57$ & $-23.53 \pm 40.67$ & $0.02$ \\
KMT-2021-BLG-0119Lb & $0.96$ & $-0.05$ & $3.13 \pm 1.11$ & $1.02 \pm 1.08$ & $-0.03 \pm 0.22$  & $0.18 \pm 0.04$ & $115.81 \pm 1.72$ & $44.87 \pm 66.61$ & $0$ & \citep{KMT-2021-BLG-0119} \\
 &  &  & $3.05 \pm 1.10$ & $1.00 \pm 1.06$ &  $-0.03 \pm 0.22$ & $0.21 \pm 0.04$ & $244.27 \pm 1.72$ & $-85.67 \pm 59.57$ & $0.11$ \\
 &  &  & $3.69 \pm 1.48$ & $1.47 \pm 1.58$ & $-0.24 \pm 0.23$  & $0.13 \pm 0.03$ & $119.81 \pm 1.66$ & $91.92 \pm 23.57$ & $5.37$ \\
  &  &  & $3.51 \pm 1.43$ & $1.40 \pm 1.51$ & $0.00 \pm 0.23$  & $0.14 \pm 0.03$ & $241.65 \pm 1.60$ & $-91.53 \pm 93.76$ & $5.67$ \\
 KMT-2021-BLG-0192Lb & $0.76$ & $0.33$ & $5.12 \pm 0.99$ & $1.64 \pm 0.37$ & $2.31 \pm 2.12$  & $0.30 \pm 0.14$  & $111.96 \pm 0.29$ & $-37.76 \pm 7.56$ & $0$ & \citep{KMT-2021-BLG-0119}\\
 &  &  & $5.26 \pm 1.01$ & $1.63 \pm 0.38$ &  $1.64 \pm 2.12$  & $0.27 \pm 0.14$ & $112.31 \pm 0.29$ & $-36.16 \pm 12.93$ & $0.83$ \\
 &  &  & $5.14 \pm 1.01$ & $2.83 \pm 0.65$ & $2.14 \pm 2.08$  & $0.29 \pm 0.14$ & $112.07 \pm 0.29$ & $-37.79 \pm 8.09$ & $0.88$ \\ 
 & $0.76$ & $0.33$ & $5.23 \pm 0.99$ & $2.76 \pm 0.63$ &  $2.90 \pm 2.14$  &$0.36 \pm 0.14$& $112.43 \pm 0.29$ & $-38.62 \pm 5.87$ & $0.90$ \\ 
KMT-2021-BLG-0171Lb & $0.77$ & $0.31$ & $4.60 \pm 1.40$ & $3.16 \pm 1.05$ & $-0.33 \pm 0.24$  & $-0.06 \pm 0.02$ & $57.54 \pm 0.63$ & $199.72 \pm 8.66$ & $0$ & \citep{KMT-2021-BLG-0171L} \\
 &  &  &  & $5.00 \pm 1.66$ & $-0.30 \pm 0.26$  & $-0.06 \pm 0.03$  & $57.77 \pm 0.57$ & $200.13 \pm 10.59$ & $0.50$ \\
 &  &  &  & $3.17 \pm 1.06$ &$-0.09 \pm 0.18$   & $-0.04 \pm 0.02$ & $57.55 \pm 0.63$ & $213.79 \pm 42.31$ & $1.70$ \\ 
 &  &  &  & $4.93 \pm 1.64$ &  $-0.07 \pm 0.19$   &$-0.04 \pm 0.02$& $57.64 \pm 0.63$ & $219.22 \pm 68.21$ & $2$ \\
 &  &  & $4.80 \pm 1.25$ & $3.76 \pm 1.02$ & $-0.15 \pm 0.25$   & $-0.45 \pm 0.02$& $59.26 \pm 0.40$ & $-101.64 \pm 28.51$ & $6.40$ \\ 
 &  & &  & $3.76 \pm 1.02$ &  $-0.16 \pm 0.18$  & $-0.04 \pm 0.02$& $59.24 \pm 0.40$ & $202.91 \pm 17.77$ & $6.80$ \\ 
 &  &  &  & $3.85 \pm 1.05$ & $-0.19 \pm 0.18$  & $-0.05 \pm 0.02$ & $59.07 \pm 0.34$ & $201.39 \pm 13.97$ & $8.60$ \\ 
 &  &  &  & $3.85 \pm 1.05$ & $-0.14 \pm 0.25$  & $-0.04 \pm 0.02$ & $59.08 \pm 0.40$ & $204.85 \pm 31.13$ & $8.80$ \\ 
MOA-2022-BLG-0249Lb & $1.03$ & $-0.17$ & $2.00 \pm 0.42$ & $1.77 \pm 0.73$ & $-0.49 \pm 0.04$  & $0.26 \pm 0.02$ & $208.44 \pm 0.12$ & $0.73 \pm 2.35$ & $0$ & \citep{MOA-2022-BLG-249L}\\
 &  &  &  & $2.00 \pm 0.75$ & $-0.56 \pm 0.04$  &$0.28 \pm 0.01$  & $206.21 \pm 0.12$ & $4.68 \pm 2.09$ & $8.10$ \\
\enddata
\end{deluxetable*}
\end{longrotatetable}

\begin{longrotatetable}
\begin{deluxetable*}{lcccccccccc}
\tabletypesize{\tiny}
\tablecaption{Additional planets whose proper motion is known from direct detection of the lense with High Contrast Imaging. The columns are the same as for Table 2 except for the parallax components, which are replaced by the proper motion components in geocentric frame.}
\tablehead{
\colhead{Event} & \colhead{$l$ (rad)} & \colhead{$b$ (rad)} & \colhead{$D_L$ (kpc)} & \colhead{$s$ (au)} & \colhead{$\mu_N$} & \colhead{$\mu_E$} & \colhead{$\alpha$ (deg)} & \colhead{$PA$ (deg)} & \colhead{$\Delta\chi^2$} & \colhead{Reference}}  
\colnumbers
\startdata 
OGLE-2005-BLG-169Lb & $0.84$ & $0.17$ & $4.00 \pm 0.40$ & $3.93 \pm 0.55$ & $4.79 \pm 0.14$ & $5.16 \pm 0.14$ & $91.91 \pm 1.72$ & $19.72 \pm 1.14$ & $0$ & \citep{OGLE-2005-BLG-169L,OGLE-2005-BLG-169L2}\\
MOA-2007-BLG-400Lb & $0.90$ & $0.06$ & $6.89 \pm 0.77$ & $6.03 \pm 0.77$ & $-2.28 \pm 0.14$ & $8.49 \pm 0.14$ & $313.48 \pm 1.72$ & $214.13 \pm 0.92$ & $0$ & \citep{MOA-2007-BLG-400L}\\
&  &  & & $0.81 \pm 0.11$ & $-2.28 \pm 0.14$ &  $8.49 \pm 0.14$& $313.30 \pm 1.95$ & $214.30 \pm 0.92$ & $0.09$ \\
MOA-2009-BLG-319Lb & $1.51$ & $-0.63$ & $7.05 \pm 0.71$ & $2.04 \pm 0.21$ & $1.79 \pm 0.14$ & $6.28 \pm 0.13$ & $150.68 \pm 0.04$ & $-42.60 \pm 1.23$ & $0$ & \citep{MOA-2009-BLG-319L,MOA-2009-BLG-319L2}\\
MOA-2013-BLG-220Lb & $0.94$ & $-0.01$ & $6.72 \pm 0.59$ & $3.03 \pm 0.27$ & $-8.23 \pm 0.11$ & $9.65 \pm 0.14$ & $278.71 \pm 0.26$ & $-87.13 \pm 0.57$ & $0$ & \citep{MOA-2013-BLG-220L} \\
\enddata
\end{deluxetable*}
\end{longrotatetable}

\bibliography{OrbitalPlanes.bbl}{}
\bibliographystyle{aasjournalv7}



\end{document}